\documentclass[usenatbib]{mn2e}
\usepackage{graphicx, amsmath, amssymb, aas_macros}
\newcommand{\vmax}{V_{\rm{max}}}
\newcommand{\rmax}{R_{\rm{max}}}

\newcommand{\mtwo}{M_{200}}
\newcommand{\mtwom}{M_{200 m}}
\newcommand{\rtwo}{R_{200}}
\newcommand{\rtwom}{R_{200 m}}

\newcommand{\dd}{{\rm d}}

\newcommand{\mstar}{M_{\star}}

\newcommand{\hmsun}{h^{-1} \, M_{\odot}}
\newcommand{\hmpc}{h^{-1} \, {\rm Mpc}}
\newcommand{\hkpc}{h^{-1} \, {\rm kpc}}
\newcommand{\millen}{MS-II}

\title[The Millennium-II Simulation]{
 Resolving Cosmic Structure Formation with the Millennium-II Simulation
}

\author[M.~Boylan-Kolchin et al.]{
 $\!\!$Michael~Boylan-Kolchin$^1$\thanks{$\!\!$e-mail: mrbk@mpa-garching.mpg.de}, 
 Volker~Springel$^1$, 
 Simon~D.~M.~White$^1$, 
 \newauthor 
 Adrian~Jenkins$^2$, and 
 Gerard~Lemson$^{3,4}$\\
 \vspace{-0.2cm}\\
 $\!\!^1$Max-Planck-Institut f\"{u}r Astrophysik, Karl-Schwarzschild-Str. 1,
 85748 Garching, Germany\\
$\!\!^2$Institute for Computational Cosmology, Department of Physics,
 University of Durham, South Road, Durham DH1 3LE, UK \\
 $\!\!^3$Astronomisches Rechen-Institut, Zentrum f\"{u}r Astronomie der
 Universit\"{a}t Heidelberg, Moenchhofstr. 12-14, 69120 Heidelberg, Germany\\
 $\!\!^4$Max-Planck-Institut f\"{u}r extraterrestrische Physik,
 Giessenbach-Str. 1, 85748 Garching, Germany}
\begin{document}

 \pagerange{\pageref{firstpage}--\pageref{lastpage}} \pubyear{2009}

\maketitle

\label{firstpage}
\begin{abstract}

  We present the Millennium-II Simulation (MS-II), a very large $N$-body
  simulation of dark matter evolution in the concordance $\Lambda$CDM
  cosmology. The MS-II assumes the same cosmological parameters and uses the
  same particle number and output data structure as the original Millennium
  Simulation (MS), but was carried out in a periodic cube one-fifth the size
  ($100 \,\hmpc$) with 5 times better spatial resolution (a Plummer equivalent
  softening of $1.0 \, \hkpc$) and with 125 times better mass resolution (a
  particle mass of $6.9\times 10^6\, \hmsun$). By comparing results at MS and
  MS-II resolution, we demonstrate excellent convergence in dark matter
  statistics such as the halo mass function, the subhalo abundance
  distribution, the mass dependence of halo formation times, the linear and
  nonlinear autocorrelations and power spectra, and halo assembly
  bias. Together, the two simulations provide precise results for such
  statistics over an unprecedented range of scales, from halos similar to those
  hosting Local Group dwarf spheroidal galaxies to halos corresponding to the
  richest galaxy clusters. The ``Milky Way'' halos of the Aquarius Project were
  selected from a lower resolution version of the MS-II and were then
  resimulated at much higher resolution. As a result, they are present in the
  MS-II along with thousands of other similar mass halos. A comparison of their
  assembly histories in the MS-II and in resimulations of 1000 times better
  resolution shows detailed agreement over a factor of 100 in mass growth. We
  publicly release halo catalogs and assembly trees for the MS-II in the same
  format within the same archive as those already released for the MS.

\end{abstract}
\begin{keywords}
 methods: $N$-body simulations -- cosmology: theory -- galaxies: halos
\end{keywords}

\section{Introduction}
In order to understand how galaxies form and evolve in their cosmological
context, we must understand the properties of dark matter halos over a wide
range of physical scales and across virtually all of cosmic history.  
Numerical simulations provide one of the best methods for approaching this
problem and have proven invaluable for studying the growth of cosmological
structure and, in particular, of dark matter halos.  Increasing computational
power and improved algorithms have led to a steady and rapid increase in the
ability of $N$-body simulations to resolve the detailed internal structure of
dark matter halos over substantial cosmological volumes.

Perhaps the most widely-used $N$-body simulation of cosmological structure
formation to date has been the Millennium Simulation \citep[hereafter
MS]{springel2005b}, which followed more than ten billion particles within a
simulation volume of $(500\,\hmpc)^{3}$.  This provided sufficient mass
resolution to see the formation of halos hosting $0.1 \,L_{\star}$ galaxies and
sufficient volume to obtain good statistical samples of rare objects such as
massive cluster halos and luminous quasars. It also enabled the implementation
of physical models for the formation and evolution of galaxy/AGN populations
throughout a large and representative cosmological volume \citep{croton2006,
  bower2006}. Since 2005, when the first results from the MS were published,
most new very large cosmological simulations have focused on larger
volumes\footnote{Recent simulations with $\ga 10^{10}$ particles within smaller
  volumes ($L_{\rm box} \approx 100 \,\hmpc$) have been used primarily for
  studying cosmic reionization at redshifts $\ga 6$ (e.g.,
  \citealt{iliev2006}).} ($L_{\rm box} \ga 1000 \,\hmpc$) in order to study
topics such as the statistical detection of baryon acoustic oscillations or
weak lensing shear, or to build mock catalogs for the next generation of galaxy
surveys \citep{fosalba2008, kim2008,  teyssier2009}.  Moving to larger volume simulations
reduces computational cost at fixed particle number both because resolved
gravitational perturbations remain linear until later times and because the
number of simulation particles in a typical nonlinear structure is smaller.

The opposite regime -- smaller volumes with higher mass resolution -- is much
more computationally demanding but is also of great interest, especially for
questions of galaxy formation, where the relevant mass scales are substantially
smaller than for large-scale clustering.  Understanding the formation and
evolution of low-mass galaxies requires adequate resolution of the dark matter
halos that host them, and this, in turn, requires {\it much} smaller particle
masses than are currently feasible for Gigaparsec-scale simulations.  These
objects are important for galaxy formation as a whole because the first
galaxies to form, which are low mass, prepare the initial conditions from which
more massive systems later form.
Another topic of particular interest
that can be addressed by high-resolution simulations is the evolution of
substructure within dark matter halos.  Such simulations show that subhalos can
lose considerable mass after being accreted onto a larger halo -- sometimes well
over 99\% -- without being completely disrupted \citep{hayashi2003, gao2004b,
  kravtsov2004}.  This means that as the resolution of a simulation is
increased, so too is the typical time between accretion of a subhalo onto a
larger system and its eventual tidal disruption.

Moving to substantially higher resolution in a large-volume simulation is
fraught with computational challenges, however.  
Increasingly small simulation time-steps are required to accurately follow
particle orbits in the dense centers of dark matter halos \citep{power2003},
where the characteristic time-scale $t_{\rm grav} \propto 1/\sqrt{G \,\rho}$
is significantly shorter than on large scales.
While only a small fraction of simulation particles reside in such dense
regions, these particles are the limiting factor in how quickly the simulation
can be evolved forward.  The maximum resolved density contrasts at $z \la 1$
can be one thousand times higher than those at $z \approx 6$; as a result,
almost all of the computational time needed for such a simulation is spent at
low redshift.  Furthermore, the strong clustering of matter within a few very
massive clumps can create serious problems with respect to parallelization: it
is much more difficult to split such a particle distribution into optimal
computational domains than is the case if the matter distribution is more
homogeneous.

In spite of these challenges, it is essential to have simulations that probe
the structure of galaxy-scale dark matter halos with high mass resolution and
over a large enough region to include a sizable and representative sample of
objects.  In this paper, we present such a calculation, the Millennium-II
Simulation (hereafter MS-II).  Section~\ref{sec:simulation} gives details of
the parameters which define this simulation and describes some of the
post-processing we have carried out on its output, in particular,
substructure-finding and merger tree-building.  We present results on the
evolution of the dark matter power spectrum and the two-point correlation
function in Section~\ref{sec:dmStatistics}.  In
Section~\ref{sec:darkMatterHalos}, we investigate the dark matter halo mass
function and the clustering bias of dark matter halos.
Section~\ref{sec:haloFormation} focuses on halo formation times, including the
dependence of clustering on formation time (so-called ``assembly bias'';
\citealt{gao2005}).  A discussion of the relation between the MS-II and the
Aquarius Project \citep{springel2008}, as well as a comparison of the assembly
histories of the halos common to the two projects, is presented in
Section~\ref{sec:aquarius}.  We summarize our results in
Section~\ref{sec:discuss}.  Throughout this paper, all logarithms without
specified bases are natural logarithms.

\section{The Millennium-II Simulation}
\label{sec:simulation}
\subsection{Simulation details}
\begin{table*}
\begin{tabular}{lcrcclcc}
 \hline
 {\bf Name}
 & $\mathbf{L_{\rm box}}$
 & \multicolumn{1}{c}{$\mathbf{N_p}$}
 & $\boldsymbol{\epsilon}$
 & $\mathbf{m_p}$
 & \multicolumn{1}{c}{$\mathbf{M_{\rm min}}$}
 & $\mathbf{M_{\rm max}}$
 & $\mathbf{f_{\rm group}}$\\
 &  [$\hmpc$] 
 & 
 & [$\hkpc$] 
 & [$\hmsun$] 
 & \multicolumn{1}{c}{[$\hmsun$]} 
 & [$\hmsun$] & \\
 \hline
 Millennium-II & 100 & 10,077,696,000 & 1.0 & $6.89 \times 10^{6}$ & $1.38 \times
 10^8$ &  $8.22 \times 10^{14}$ & 0.601\\
 Millennium & 500 & 10,077,696,000 & 5.0  & $8.61\times 10^{8}$ & $1.72 \times
 10^{10}$ & $3.77 \times 10^{15}$ & 0.496\\
 mini-MS-II & 100 & 80,621,568 & 5.0 &  $8.61\times 10^{8}$ & $1.72 \times
 10^{10}$ &  $8.29 \times 10^{14}$ & 0.502\\
\hline
\end{tabular}
\caption{
 Some basic properties of the new Millennium-II Simulation are compared to
 those of the MS and to the
 lower resolution version of MS-II (mini-MS-II).  $L_{\rm box}$
 is the side length of the simulation box, $N_p$ is the
 total number of simulation particles used, and $\epsilon$ is
 the Plummer-equivalent force softening of the simulation, in comoving units.
 $m_p$ gives the mass of each simulation particle, $M_{\rm min}$ gives
 the mass of the smallest FOF halos (corresponding to our choice of storing
 all halos with $N_p \ge 20$), and
 $M_{\rm max}$ gives the maximum FOF halo mass found in the simulation.  $f_{\rm
   group}$ is the fraction of all simulation particles in FOF groups of
 20 or more particles at $z=0$.
 \label{table:comparison}}
\end{table*}
\label{sec:simulationDetails}
The Millennium-II Simulation follows $2160^3$ particles within a cubic
simulation box of side length $L_{\rm box}=100 \,h^{-1} \,{\rm Mpc}$. This is
five times smaller than $L_{\rm box}$ for the Millennium Simulation.  The
volume sampled by the \millen\ is thus 125 times smaller than in the MS but the
mass resolution is correspondingly 125 times better: each simulation particle
has mass $6.885\times 10^{6}\,\hmsun$.  With this mass resolution, halos
similar to those hosting Local Group dwarf spheroidals are resolved at our 20
particle mass limit, while halos of Milky Way-mass galaxies have hundreds of
thousands of particles and halos of rich clusters have over fifty million
particles.  The Plummer-equivalent force softening\footnote{see eqn. 4 of
  \citet{springel2005} and corresponding text for details of how force
  softening is implemented in {\tt GADGET}.} adopted for the \millen\ was $1 \,
h^{-1} \, {\rm kpc}$ and was kept constant in comoving units; this value
corresponds to 0.06\% (10\%) of the virial radius for the largest (smallest)
halos at redshift zero.

The $\Lambda$CDM cosmology used for the \millen\ is identical to
that of the MS and the Aquarius simulations:
\begin{eqnarray}
 \label{eq:cosmo_params}
 & & \Omega_{\rm tot} = \! 1.0, \; \Omega_m = \!0.25, \; \Omega_b=0.045, \; \Omega_{\Lambda}=0.75, \nonumber \\
 & & h = 0.73, \;\sigma_8=0.9, \; n_s=1\,,
\end{eqnarray}
where $h$ is the Hubble constant at redshift zero in units of 100 km s$^{-1}$
Mpc$^{-1}$, $\sigma_8$ is the rms amplitude of linear mass fluctuations in $8
\,\hmpc$ spheres at $z=0$, and $n_s$ is the spectral index of the primordial
power spectrum.  Retaining the cosmological parameters of the MS allows us to
test for convergence by comparing results in the regime where objects are
well-resolved in both simulations as well as to extend the range of structures
probed by combining, when appropriate, results from the two simulations.  In
particular, this helps us understand the effects of resolution at low particle
number.

The initial conditions for the simulation were created at redshift $z=127$,
identical to the starting redshift of the MS, using a ``glass'' initial
particle load \citep{white1996}; the initial particle positions and velocities
were then computed using the displacement field tabulated on a $4096^3$ mesh
and the Zeldovich approximation.  The transfer function used for calculating
the input linear power spectrum was computed with the Boltzmann code {\tt
  CMBFAST} \citep{seljak1996}.

The amplitudes and phases of the initial linear fluctuation modes in the \millen\
are identical to those in the simulation from which the Aquarius Project halos
were chosen.  Specifically, all modes with a wave-vector whose maximum
component is less than $13.57 \, h \, {\rm Mpc}^{-1}$ have amplitudes and
phases that match those of the Aquarius simulations; all other modes were set
at random to have the same underlying power spectrum. The Aquarius halos are
thus present in the \millen.  A discussion of the relationship between the
Aquarius Project and the \millen\ is presented in Section~\ref{sec:aquarius}.

The \millen\ was run with {\tt GADGET-3}, an updated version of the {\tt
  GADGET} code \citep{springel2001, springel2005}.  {\tt GADGET-3} is a TreePM
code: long-range force calculations are performed with a particle-mesh
algorithm while short-range forces are calculated via a hierarchical tree.
While the original MS was performed with a memory-optimized version of {\tt
  GADGET-2}, the extremely high level of clustering that occurs in the \millen\
results in somewhat different computational requirements; in particular, a more
flexible domain decomposition is necessary. {\tt GADGET-3} was developed
specifically for this situation.

The \millen\ was performed on the IBM Power-6 computer at the Max-Planck
Computing Center in Garching, Germany, using 2048 cores and approximately 8 TB
of memory.  A Fast Fourier Transform with $4096^3$ cells was used for the PM
calculation.  Particles were allowed to have individual, adaptive time-steps.
The evolution of the simulation required approximately 1.4 million CPU hours
and $2.77\times 10^{13}$ force calculations for the 22,142 simulation
time-steps.  The mid-point of the simulation in terms of computational time was
$z=0.88$; by contrast, evolving the simulation from $z=127$ to $z=6$ took only
10\% of the total CPU time.

Outputs were saved at sixty-eight epochs: sixty-five snapshots spaced according
to
\begin{equation}
 \label{eq:outputs}
 \log_{10}(1+z_N) = \frac{N \, (N+35)}{4200} \quad (0 \le N \le 64)
\end{equation}
and three high redshift outputs at $z=40, \, 80, \, {\rm and} \, 127\,$.  The
spacing scheme in Equation~(\ref{eq:outputs}) is identical to that used in the
MS.  We have extended the range of regularly-spaced outputs
to $z \approx 31.3$, however, because the increased mass resolution of
the \millen\ results in earlier-forming first structures. 

For comparison purposes, we have also performed a version of the \millen\ with
identical initial conditions and in the same volume with the same outputs as
the main run but at the same mass and force resolution as the original MS (so
$N_p=432^3$).  This ``mini-MS-II'' simulation allows us to test how numerical
resolution affects our results.  Some basic details of all three simulations
are listed in Table~\ref{table:comparison}.
\begin{figure*}
 \centering
 \vspace{2cm}
 \includegraphics[scale=0.375]{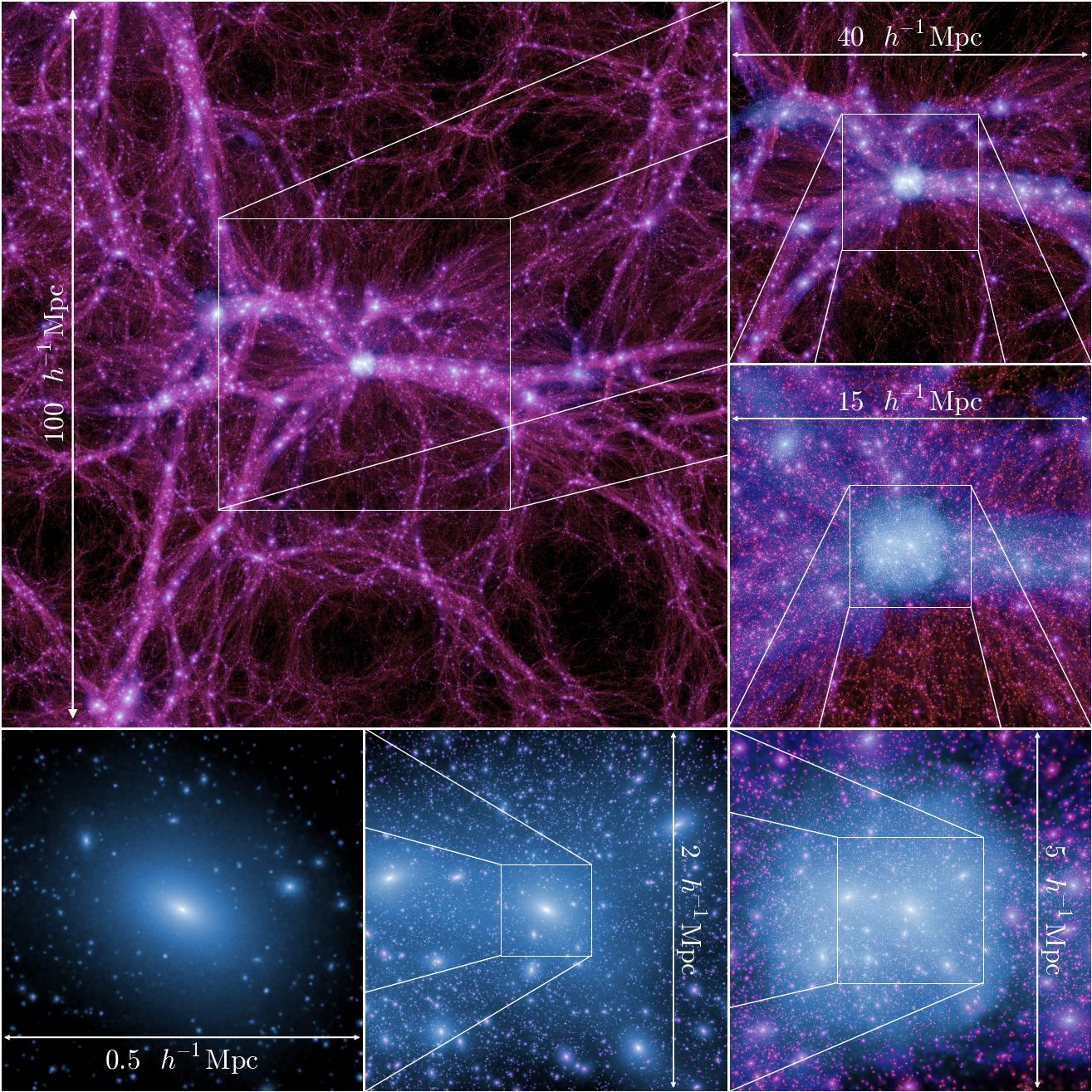}
 \caption{
   \label{fig:zoom_panels} 
   A sequential zoom through the Millennium-II Simulation.  The large image
   (upper left) is a $15 \,\hmpc$ thick slice through the full $100\, \hmpc$
   simulation box at redshift zero, centered on the most massive halo in the
   simulation.  This FOF halo has $M_{\rm FOF}=8.2 \times 10^{14} \, \hmsun$,
   similar to the mass of the Coma cluster \citep{colless1996}, is composed of
   119.5 million particles, and contains approximately 36,000 resolved subhalos
   spanning 6.7 decades in mass.  Starting from the upper right and moving
   clockwise, subsequent panels zoom into the cluster region and show slices
   that are 40, 15, 5, 2, and 0.5 $\hmpc$ on a side (with thicknesses of 10, 6,
   5, 2, and 0.5 $\hmpc$).  Even at $0.5 \, \hmpc$, which is approximately
   $1/10^{\rm th}$ the diameter of the halo, a rich variety of substructure is
   visible.  }
\end{figure*}

\begin{figure*}
 \centering
 \includegraphics[scale=0.35]{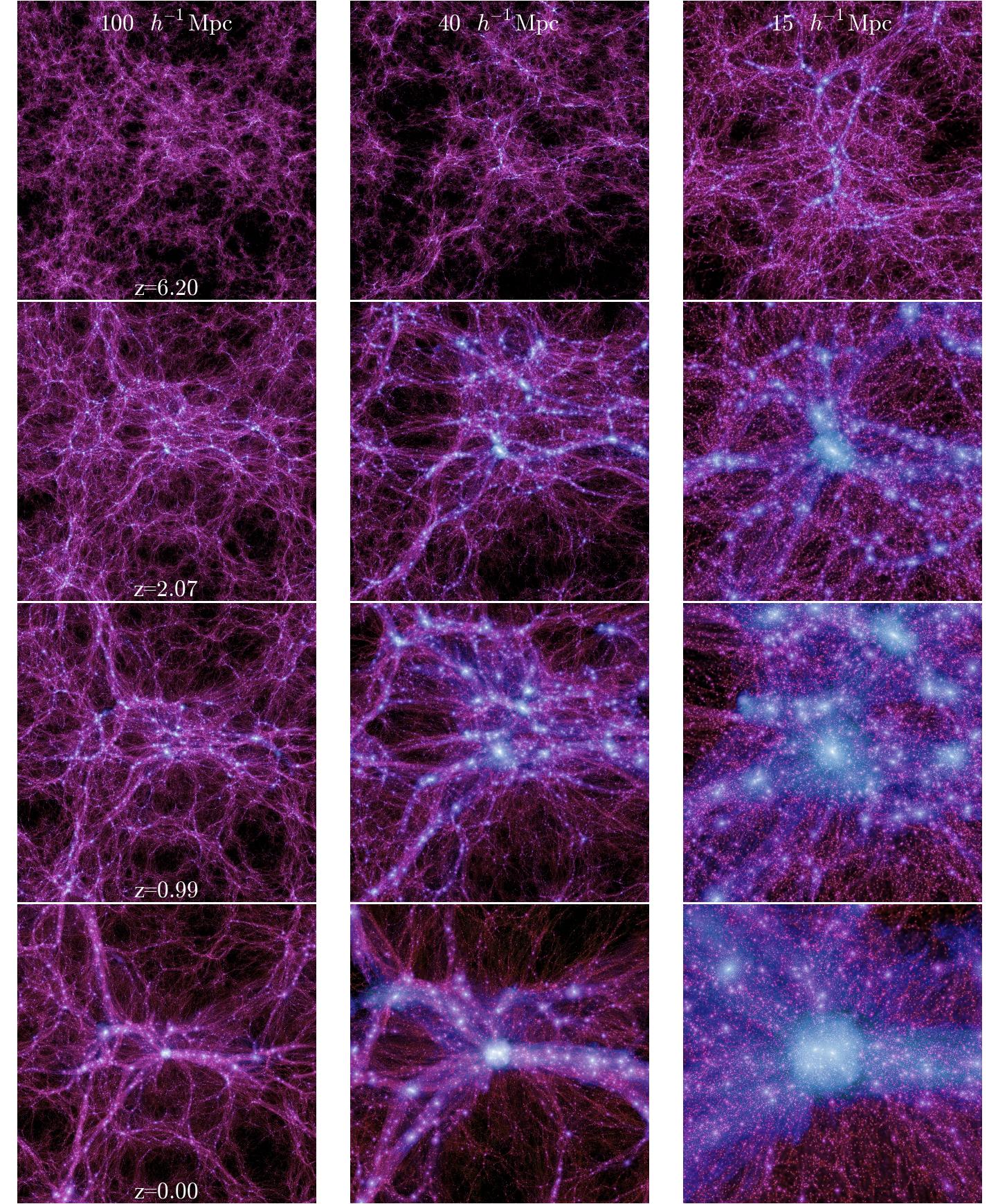}
 \caption{
   \label{fig:evol_panels} 
   Time evolution of the largest FOF halo at $z=0$ in the Millennium-II
   Simulation.  The halo is shown at three co-moving scales (from left to right:
   100, 40, and 15 $\hmpc$, with thickness 15, 10, and 6 $\hmpc$) and at four
   different cosmological epochs (from top to bottom: $z$=6.2, 2.07, 0.99, and
   0).}
\end{figure*}

\subsection{Halos and Subhalos}
\label{sec:halosAndSubhalos}
Dark matter halos were identified on-the-fly during the simulation for each
snapshot using the friends-of-friends (FOF) algorithm \citep{davis1985} with a
linking length of $b=0.2$; all groups with at least 20 particles were retained
for later analysis.  This process resulted in $1.17 \times 10^7$ FOF groups at
$z=0$, slightly fewer than the peak value of $1.53 \times 10^7$ at $z=3.06$.
Just over 60\% of the particles in the full simulation belong to a FOF group at
$z=0.$ A catalog with quantities of interest for each FOF halo (e.g., position,
velocity, number of particles), as well as a list of the particles composing
each halo, was saved at each snapshot.

The largest FOF group at $z=0$, a cluster-mass dark matter halo, has over 119
million particles.  Figure~\ref{fig:zoom_panels} shows images of the dark
matter distribution in the \millen\ on a number of different physical scales,
all centered on this halo\footnote{Images of individual panels and additional
  information related to the \millen\ are available at \\
  http://www.mpa-garching.mpg.de/galform/millennium-II}.  The large panel in the
upper left shows a $15 \,\hmpc$-thick slice through the full simulation volume
($100 \, \hmpc$ on each side).  The well-known cosmic web of filaments and
voids can be seen clearly.  Starting in upper-right and moving clockwise, the
other five panels zoom successively closer into the halo.  The bottom-right
panel is $5 \,\hmpc$ on a side, approximately the diameter of the halo.  As has
been long known (e.g., \citealt{moore1998, tormen1998, ghigna1998, klypin1999a,
  klypin1999, moore1999}), FOF halos in $\Lambda$CDM simulations are not
monolithic objects but rather are teeming with substructure; this substructure
is clearly evident even at 1/10$^{\rm th}$ the radius of the halo
(lower-left panel).

During post-processing, every FOF halo was searched for bound dark matter
substructure using the {\tt SUBFIND} algorithm \citep{springel2001a}.  {\tt
  SUBFIND} identifies substructures within a FOF halo by searching for
overdense regions using a local SPH density estimate, identifying substructure
candidates as regions bounded by an isodensity surface that traverses a saddle
point of the density field, and testing that these potential substructures are
physically bound with an iterative unbinding procedure.  All self-bound
structures with at least 20 particles were deemed to be physical {\it subhalos}
and were stored in subhalo catalogs.  Several properties of each subhalo were
also tabulated and saved, including velocity dispersion, peak circular velocity
$\vmax$ and the radius $\rmax$ at which $\vmax$ is attained, half-mass radius,
spin, position, and velocity.  The member particles of each subhalo were ranked
according to binding energy and stored in that order, which facilitates
tracking subhalos across simulation outputs.  Note that with these procedures,
we have two separate but related sets of dark matter structures: FOF halos and
subhalos.

While each subhalo has a single well-defined mass assigned to it -- the sum of
the masses of its constituent particles -- multiple mass definitions for FOF
halos are common in the literature (see \citealt{white2001} for a discussion of
subtleties associated with assigning masses to halos).  The most
straightforward definition is $M_{\rm FOF}$, the total mass of all the member
particles. Another possibility is $M_{\Delta}$, defined as the mass contained
in a spherical region (centered on the particle in the dominant subhalo with
the minimum gravitational potential) with average density a factor $\Delta$
larger than the critical density of the universe.  For each FOF halo, we
calculated $M_{\Delta}$ for $\Delta=200$, $200 \, \Omega_m(z)$, and
$\Delta_v(z)$, where the last value is taken from the spherical top-hat
collapse model (see, e.g., \citealt{bryan1998}).  We refer to spherical
overdensity masses as $\mtwo$ [$\Delta=200$], $\mtwom$ [$\Delta=200 \,
\Omega_m(z)$], or $M_v$ [$\Delta=\Delta_v(z)$] and to the corresponding virial
radii as $\rtwo$, $\rtwom$, or $R_v$.  At high redshifts, when the matter
density is nearly equal to the critical density, all three definitions give
similar masses.  At lower redshifts, $\mtwom >M_v>\mtwo$ for a given halo.

\subsection{Merger Trees}
\label{sec:mergerTrees}
Merger trees were constructed at the subhalo level by requiring subhalos to
have at most one descendant.  For many subhalos, this descendant can be found
trivially (if it exists): all particles in a subhalo at snapshot $S_n$ may
belong to a single subhalo at the subsequent snapshot $S_{n+1}$, in which case
this subhalo is clearly the descendant of the subhalo at the previous snapshot.
There is also the possibility that particles belonging to one subhalo at $S_n$
may be distributed over more than one subhalo at $S_{n+1}$.  We still require
each subhalo to have at most one descendant for these cases, so a
subhalo's unique descendant is identified as follows.  First, the binding energy
of each particle in the subhalo at $S_n$ is calculated and the particles are
ranked by this binding energy.  Each potential descendant subhalo -- that is,
each subhalo at $S_{n+1}$ containing at least one particle $j$ from the subhalo at
$S_n$ -- is then given a score $\chi$ that is based on the binding energy rank
$\mathcal{R}$ of these particles: $\chi = \sum_j \mathcal{R}_j^{-2/3}$.  The
subhalo at $S_{n+1}$ with the largest value of $\chi$ is defined to be the
descendant subhalo.  This procedure weights the most bound regions of a
 subhalo most heavily when determining its descendant.  Note that while
descendants are unique, a given subhalo may have many progenitors.

There is one slight complication to this process.  Sometimes a subhalo passing
through the dense center of a larger system will not be identified by {\tt
  SUBFIND}, simply because the density contrast is not high enough.  To
mitigate this problem, we also search for a descendant at snapshot $S_{n+2}$.
In the vast majority of cases, however, the descendant of a subhalo is found at
$S_{n+1}$.

Once all unique descendants are found, the subhalos are linked across all
snapshots to form merger trees.  This is done by taking a subhalo at $z=0$ and
linking all subhalos with descendant pointers to this halo, then repeating with
all of those subhalos, and so on, until no more subhalos can be joined.  This
process results in links between most, though not all, of the subhalos in the
simulation: subhalos that are never connected to any $z=0$ subhalo and that are
never connected to any progenitor of any $z=0$ subhalo are not included in the
trees.  We save several pointers for each tree subhalo for later use.  These
include pointers to the dominant subhalo of the subhalo's FOF group, the next
most massive subhalo in the FOF group, the progenitor that contains the largest
fraction of the subhalo's particles, the subhalo's descendant, and the next
most massive subhalo that shares the same descendant\footnote{see also figure
  5 in the Supplementary Information of \citet{springel2005b}}.

The merger trees for the \millen\ contain approximately 590 million subhalos in
total (as compared to 760 million subhalos in the MS).  While the overall data
volume of the \millen\ is similar to that of the MS ($\approx 25$ Terabytes,
dominated by the raw particle data), the highly clustered nature of the \millen\
means that the trees are markedly less homogeneous.  There are only half as many
trees in total in the \millen\ as in the MS, but the largest tree is {\it much}
larger, with over 90 million subhalos (compared to 500 thousand for the largest
tree in the MS).

\subsection{An example of subhalo tracking}
As an example of our merger trees and subhalo tracking, we consider
the main progenitor histories of the most massive halo at $z=0$ and of
two of its subhalos.  We use both the \millen\ and mini-MS-II in order to
highlight the effects of resolution and to probe the convergence of
the subhalo identification and merger tree building algorithms.
Objects can be matched between the simulations because the initial
conditions are identical on all scales that overlap; any differences
are due to force and mass resolution and to differing discreteness
effects.

The main cluster halo is trivial to find in both simulations.  At $z=0$, the
properties agree quite well between the two: the FOF masses are the same to
within 1\% ($M_{\rm FOF}=8.22\times 10^{15} \,\hmsun$ for \millen\ versus
$8.29\times 10^{15} \,\hmsun$ for mini-MS-II) and even the position of the
halo, as determined by the gravitational potential minimum, agrees to with
$0.022 \,\hmpc=4\,\epsilon$ (for mini-MS-II).  Figure~\ref{fig:evol_panels}
shows the main progenitor of this halo on three co-moving scales (from left to
right: 100, 40, and 15 $\hmpc$) and at four redshifts (from top to bottom:
$z=6.2$, 2.07, 0.99, and 0.0).
\label{sec:mergerTreeExample}
\begin{figure}
 \centering
 \includegraphics[scale=0.58, viewport=0 0 421 405, clip]{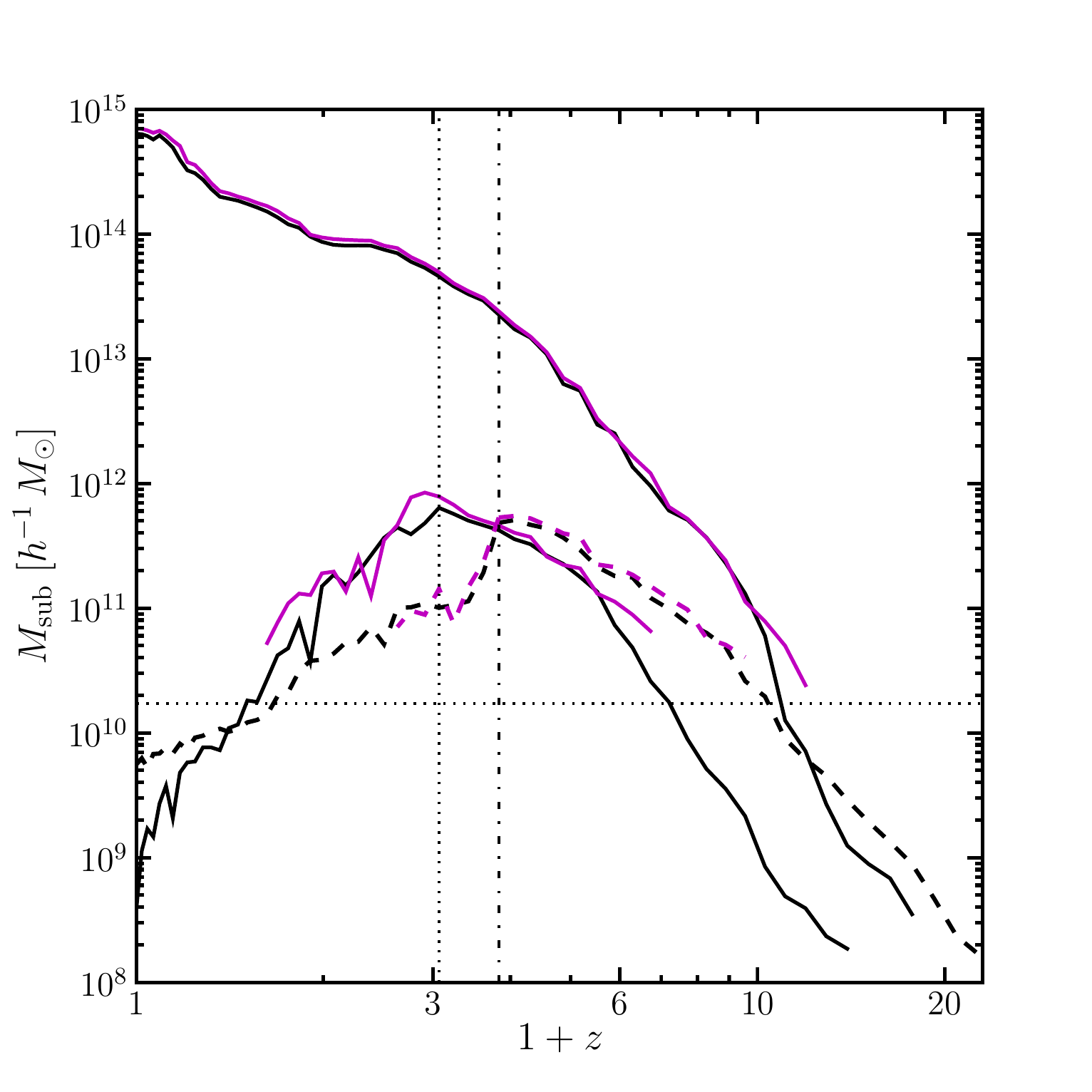}
 \caption{
   \label{fig:subHistory}
   Subhalo assembly histories in the \millen\ and mini-MS-II.  The upper set of
   curves shows the mass in the main progenitor branch of the most massive
   $z=0$ FOF group for the \millen\ (black) and mini-MS-II (magenta), while the
   lower set of curves shows main progenitor histories for subhalos that are
   accreted onto the FOF group of the cluster at $z \approx 2$ (solid lines)
   and $z \approx 3$ (dashed lines).  The horizontal dotted line shows the
   mass resolution of MS and mini-MS-II, while the vertical dotted and
   dot-dashed lines show the epochs at which the smaller halos joined the FOF
   group of the main cluster in the \millen.}
\end{figure}
 Using the merger trees, we track the main progenitor of the most
massive cluster back in time for each run until there are no further
progenitors.  The top two curves in Figure~\ref{fig:subHistory} show
the mass of the central subhalo of this main progenitor
branch\footnote{Note that the while the FOF masses of the halos agree
 to within 1\% between the \millen\ and mini-MS-II, the mass of the
 central subhalo in the \millen\ is slightly smaller because more
 distinct subhalos are identifiable.  Many subhalos that are
 resolvable in the \millen\ but not in mini-MS-II show up as extra mass
 in the central subhalo in mini-MS-II.}  for the \millen\ (black) and
mini-MS-II (magenta).  The two are in excellent agreement from $z=0$
to $z=9$, at which point the main branch from mini-MS-II falls below
100 particles and resolution effects become relevant (the main
progenitor in the \millen\ can be traced back all the way to $z \approx
18$).  Clearly, the assembly of the main progenitor is very well
converged between the two runs at $z \la 9$.

We also consider the evolution of two far less massive subhalos within this FOF
group: subhalo A (lower solid lines in Figure ~\ref{fig:subHistory}) and
subhalo B (dashed lines) are identified in the \millen\ at $z=0$.  They are not
massive enough to be identified at $z=0$ in mini-MS-II, as subhalo A has 57
particles and subhalo B has 812 at the final snapshot (the resolution limit of
mini-MS-II corresponds to 2500 particles from the \millen\ and is marked by
the horizontal dotted line).  Nevertheless, Figure~\ref{fig:subHistory} shows
that by tracking subhalos A and B backward (lower black curves), we find that
both were much more massive in the past: each exceeded $5 \times 10^{11}
\,\hmsun$ at one point in its history (92,431 particles for A and 73,718 for B)
and each was the central subhalo of its own FOF group before falling into the
main progenitor of the cluster (this accretion happened at $z \approx 2$ for
subhalo A, marked by the vertical dotted line, and at $z \approx 3$ for subhalo
B, marked by the vertical dot-dashed line)\footnote{A and B are two of four
  subhalos in the main FOF cluster at $z=0$ that (i) have $N_p(z=0) < 1000$ and
  (ii) have over 1500 progenitor subhalos in their sub-tree.  This selection
  picks out halos that were massive at one point in their history but are not
  at $z=0$.  We have also investigated the other two subhalos from this sample
  and find similar convergence in the subhalo tracking and mass identification
  between the two simulations; for clarity, the results are not plotted in
  Figure~\ref{fig:subHistory}.}.

These maximum masses for A and B are easily resolvable in mini-MS-II, so we
can hope to find the equivalent subhalos there and compare their mass
histories.  It is in general unrealistic to expect subhalos to have identical
positions in runs of differing resolution: the gravitational force is softened
at different scales and there is a difference in the `graininess' of the
gravitational potential due to finite particle mass, both of which can cause
orbital phase offsets that accumulate over time\footnote{ See
  \citet{springel2008} for a method to match subhalos in simulations based on
  the positions of the subhalo particles in the initial
  conditions.}. Nevertheless, we are able to locate A and B in mini-MS-II and
we find them to be at almost exactly the same positions as in the \millen: the
absolute positions for subhalo A (subhalo B) differ by only 0.015 (0.010)
$\hmpc$ at the times marked by the vertical lines, which is only three times
the force softening of the low resolution run.  
\begin{figure}
 \centering
 \includegraphics[scale=0.58, viewport=0 0 421 405, clip]{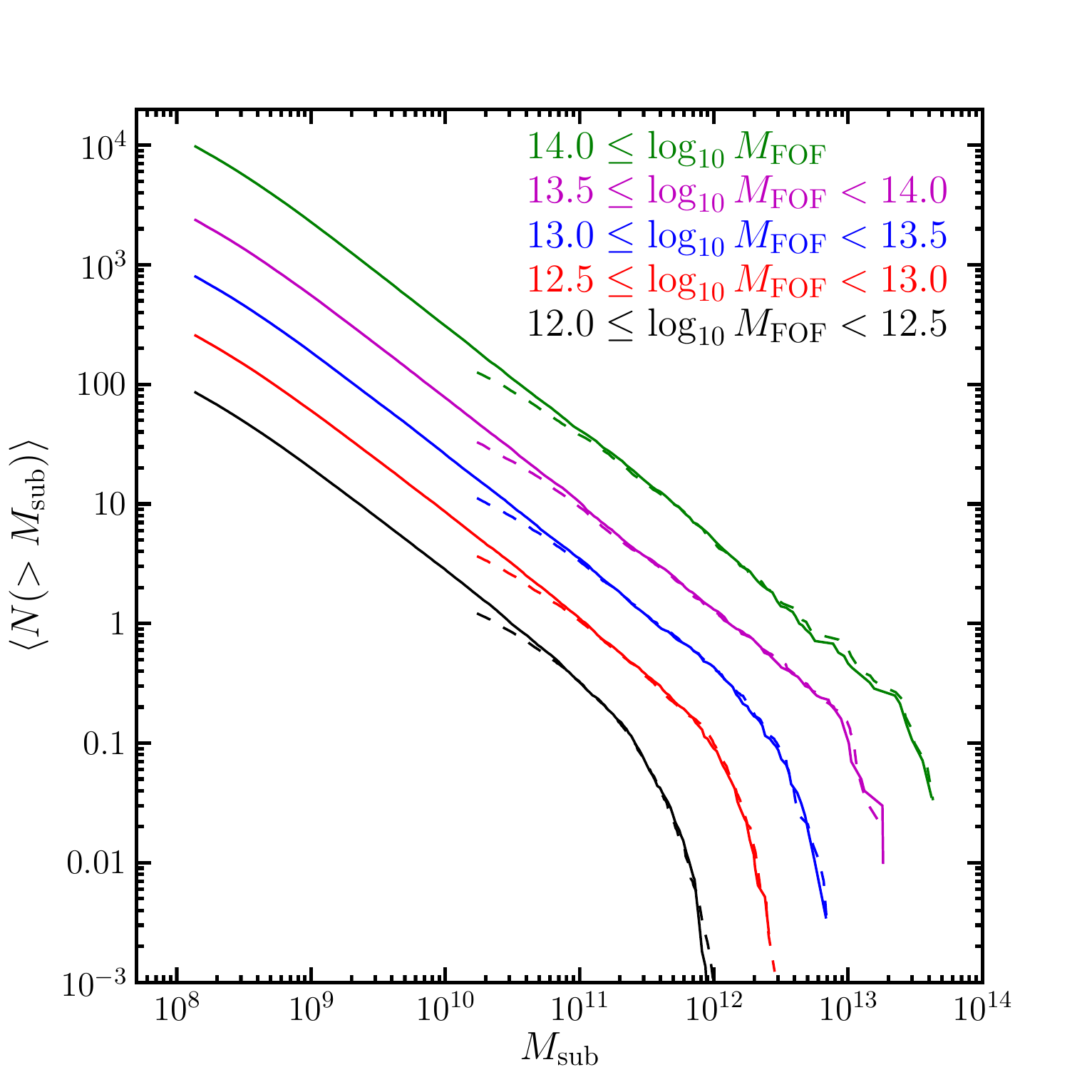}
 \caption{
   \label{fig:subsConverge}
   The mean cumulative subhalo abundance per parent halo in the \millen\ (solid
   curves) and mini-MS-II (dashed curves) in five parent halo mass bins.  The
   curves for each simulation are plotted down to the 20 particle resolution
   limit.  There is a deficit in subhalos at this limit in mini-MS-II
   relative to the \millen\ due to resolution.  At 3-5 times the minimum
   resolution limit, however, the two simulations agree very well, indicating
   that subhalos with more than 100 particles are reliably resolved.  }
\end{figure}

Having located A and B in mini-MS-II, we then use the merger trees to track the
subhalos both forward and backward in time and we compare to the results from
the \millen. These subhalo mass histories are shown in the lower magenta lines
(solid for subhalo A, dashed for subhalo B) of Figure~\ref{fig:subHistory}.  We
see that the subhalos in the two simulations have remarkably similar assembly
histories, not only when they are the main FOF subhalo (to the right of the
vertical lines) but also when they are non-dominant subhalos within a larger
FOF group (to the left of the vertical lines).  This regime, where the subhalos
are subjected to strong tides that vary rapidly in time, can be extremely
difficult to capture accurately in simulations of differing resolution.  The
excellent agreement between the lower magenta and black curves demonstrates
that the subhalos have the same dynamical histories, are assigned the same
masses, and are linked in the same way by the merger trees in the two runs.

Figure~\ref{fig:subHistory} also illustrates resolution limitations. As tides
strip mass from the subhalos, they are lost from the mini-MS-II catalog and are
considered to have merged with the dominant subhalo, while they persist as
independent subhalos to $z=0$ at the significantly enhanced resolution of
the \millen.  The high resolution of the \millen\ is required to study the fates of
subhalos hosting low mass galaxies within larger structures: note that the
maximum masses -- approximately $6 \times 10^{11} \,\hmsun$ -- of A and B are
quite large, larger than the halo masses of likely Milky Way progenitors at the
redshift of accretion into the massive halo.  A and B are therefore likely to
host galaxies of stellar mass comparable to that of the Milky Way. At
mini-MS-II resolution (i.e. MS resolution) it is not possible to follow the
dynamics of these subhalos past $z \approx 1.5$ for subhalo B or $z \approx
0.5$ for subhalo A, by which redshifts their masses have dropped below $10^{11}
\,\hmsun$.  The \millen\ captures the later dynamical history of both subhalos,
even though subhalo A (subhalo B) retains only 0.06\% (1.1\%) of its maximum
mass at $z=0$. Note that these final masses are considerably smaller than the
likely {\it stellar} masses of the associated galaxies, so it remains unclear
how realistic the late-time dynamical evolution actually is in MS-II.

We can also consider the statistical agreement between the \millen\ and mini-MS-II
by comparing stacked subhalo abundances as a function of host halo mass.
Figure~\ref{fig:subsConverge} shows the mean number of subhalos per host halo
in five host halo mass bins for the two simulations.  Resolution effects reduce
the number of subhalos at a given subhalo mass in mini-MS-II (dashed curves)
relative to the \millen\ (solid curves) for subhalos with few particles: at the
minimum resolvable mass of 20 particles in mini-MS-II, the abundance of
subhalos is reduced by approximately 30\% relative to the \millen.  The results
from the two simulations are in excellent agreement for more massive subhalos
($M_{\rm sub} > 10^{11} \,\hmsun$), showing that subhalos containing at least
50-100 particles are reliably resolved.

\section{Statistics of the Density Field}
\label{sec:dmStatistics}
\subsection{Power Spectrum}
\begin{figure}
  \centering
   \includegraphics[scale=0.58, viewport=0 0 421 464, clip]{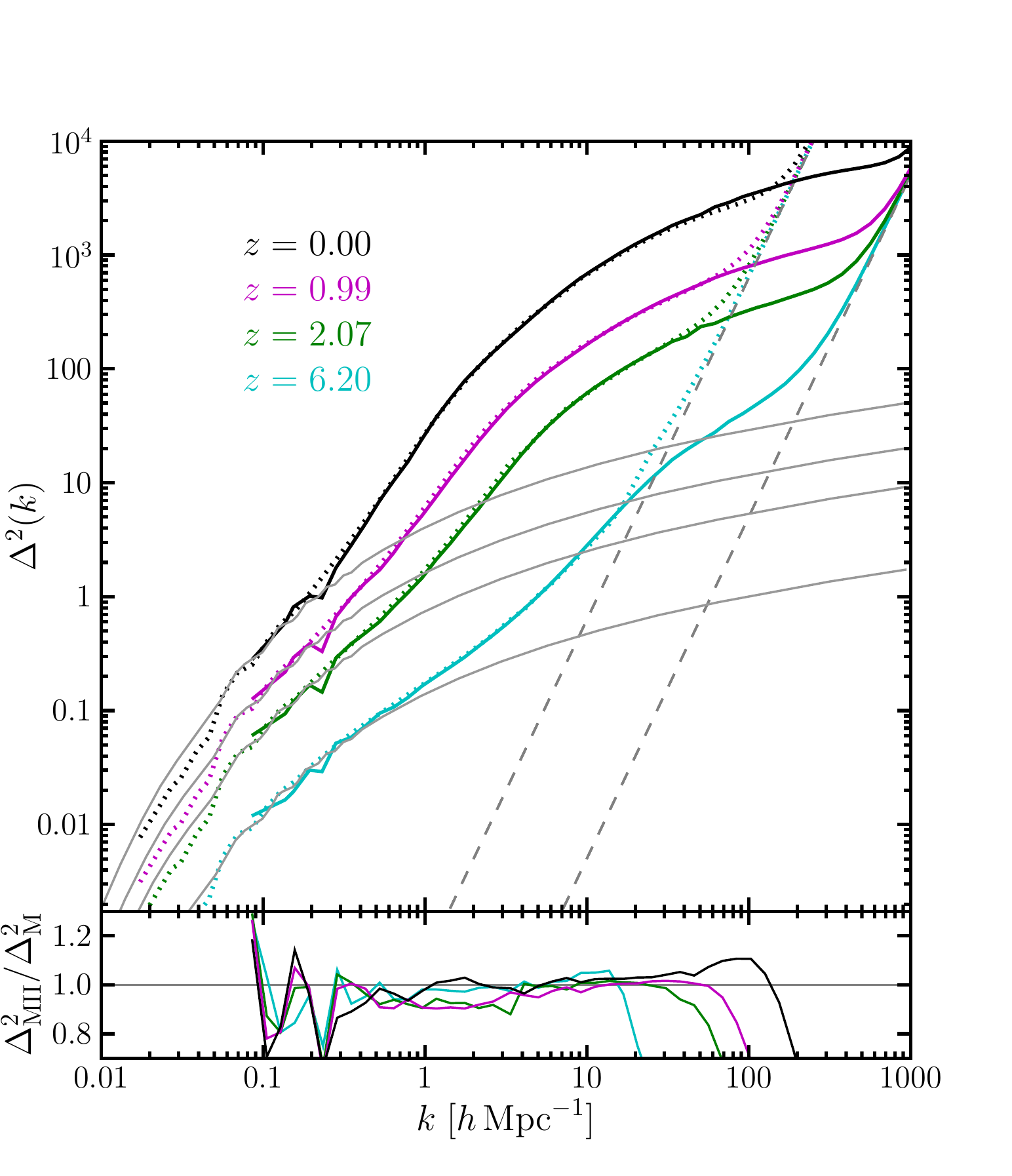}
   \caption{
     \label{fig:powerspec}
     The power spectrum $\Delta^2(k)$ measured from the \millen\ at redshifts
     0.0 (black curves), 0.99 (magenta), 2.07 (green), and 6.20 (cyan), as well
     as the linear theory power spectrum at each redshift (gray curves).  Power
     spectra from the MS (dotted curves) at the same redshifts are also shown
     for comparison.  The dashed lines correspond to the shot noise limit for
     the \millen\ (right) and the MS (left); the power spectra have not been corrected
     for shot noise.  The bottom panel shows the ratio of the power spectra.}
\end{figure}
\begin{figure}
 \centering
 \includegraphics[scale=0.58, viewport=0 0 421 464, clip]{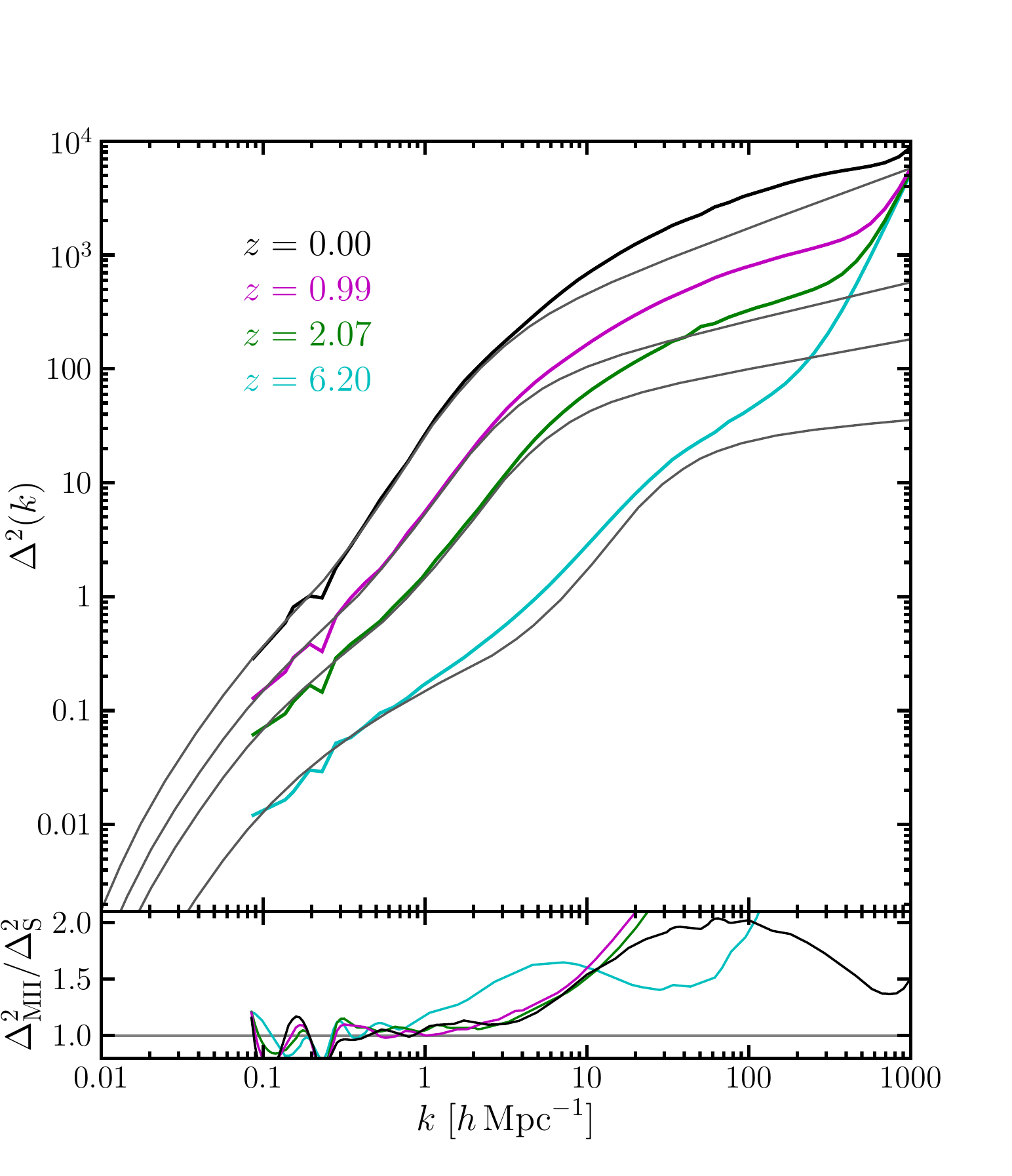}
   \caption{
     \label{fig:smithPk}
     A comparison of the \millen\ power spectra at four redshifts to the halo
     model fit from Smith et al. (\citeyear{smith2003}; gray curves).  The fit
     at $z=0$ is accurate for $k \la 7 \, h \,{\rm Mpc}^{-1}$ but
     underestimates the power spectrum from the \millen\ by 50\% or more for
     $k$ between 10 and 100 $h$ Mpc$^{-1}$.  Redshifts one and two show similar
     results but the agreement at $z=6$ is poor.}
\end{figure}
At comoving position $\mathbf{x}$ and time $t$, the mass density field can be
expressed as 
\begin{equation}
 \label{eq:delta}
 \rho(\mathbf{x}, t)= \bar{\rho}(t) \,[1+\delta(\mathbf{x},t)] \,.
\end{equation}
In the standard picture of structure formation in a cold dark matter universe,
the initial density fluctuation field $\delta(x, 0)$ is taken to be a Gaussian
random field. Its statistical properties are therefore fully specified
by its power spectrum $P(k)$
or equivalently its dimensionless power spectrum $\Delta^2(k)$,
\begin{equation}
 \label{eq:delta2}
 \Delta^2(k) \equiv \frac{k^3} {2 \pi^2}\, P(k) \,.
\end{equation}
$\Delta^2(k)$ measures the power per logarithmic interval in wavenumber;
$\Delta^2(k) \approx 1$ therefore indicates that fluctuations in density on
scales near wavenumber $k$ are of order unity.

The dark matter power spectrum from the \millen\ is shown in
Figure~\ref{fig:powerspec}.  We plot the results at four redshifts: $z=0$
(black curves), 0.99 (magenta), 2.07 (green), and 6.20 (cyan).  On large
physical scales (small wavenumber $k$), the power spectrum follows the
prediction from linear theory (light gray lines).  As time progresses, larger
and larger physical scales become non-linear and the small-scale power exceeds
the linear theory prediction.  Results from the MS are also included in
Figure~\ref{fig:powerspec} for comparison (dotted curves).  The agreement
between the two simulations is very good, and the \millen\ extends the
measurement of $\Delta^2(k)$ by a factor of 5 at large $k$.  We have not
performed a shot noise correction in this figure, as it not clear that it is
appropriate to do so (see, e.g., \citealt{baugh1995, sirko2005}). The shot
noise limit for each run is plotted as a dashed gray line.

The large dynamic range and uniform mass resolution of the \millen\ allows us to
probe the dark matter power spectrum on a wide range of scales, including scales
where existing fitting functions \citep{peacock1996, smith2003} are
uncalibrated and untested.  Figure~\ref{fig:smithPk} compares the power
spectrum computed from the \millen\ with the halo model predictions of Smith et
al. (gray lines).  At redshift 2 and below, the Smith et al. model agrees with
the calculated power spectrum to within 10\% for $k < 5 \, h \,\rm{Mpc}^{-1}$.
At larger $k$, however, the model significantly underestimates the power, with
errors exceeding a factor of 2.  The Smith et al. model was not calibrated for
this range, so it is not surprising that it fails to reproduce the simulation
results.  Nevertheless, this is a reminder that extrapolating fitting functions
beyond their calibrated range can lead to serious errors.  At $z=6$, the Smith
et al. model fails to fit the data over the range $1 < k < 10 \, h \,{\rm
  Mpc}^{-1}$, where the \millen\ and MS agree very well.

At scales of $k \ga 10 \,h\, \rm{Mpc}^{-1}$, baryonic physics plays an
important role in determining the real dark matter power spectrum
(e.g., \citealt{rudd2008}).  Although a full modeling of baryonic
effects will be necessary to get accurate predictions for $\Delta^2$
at these scales, understanding the underlying dark matter-only power
spectrum still provides a critical baseline.

\subsection{Two Point Correlation Function}
\begin{figure}
 \centering
 \includegraphics[scale=0.58, viewport=0 0 421 464, clip]{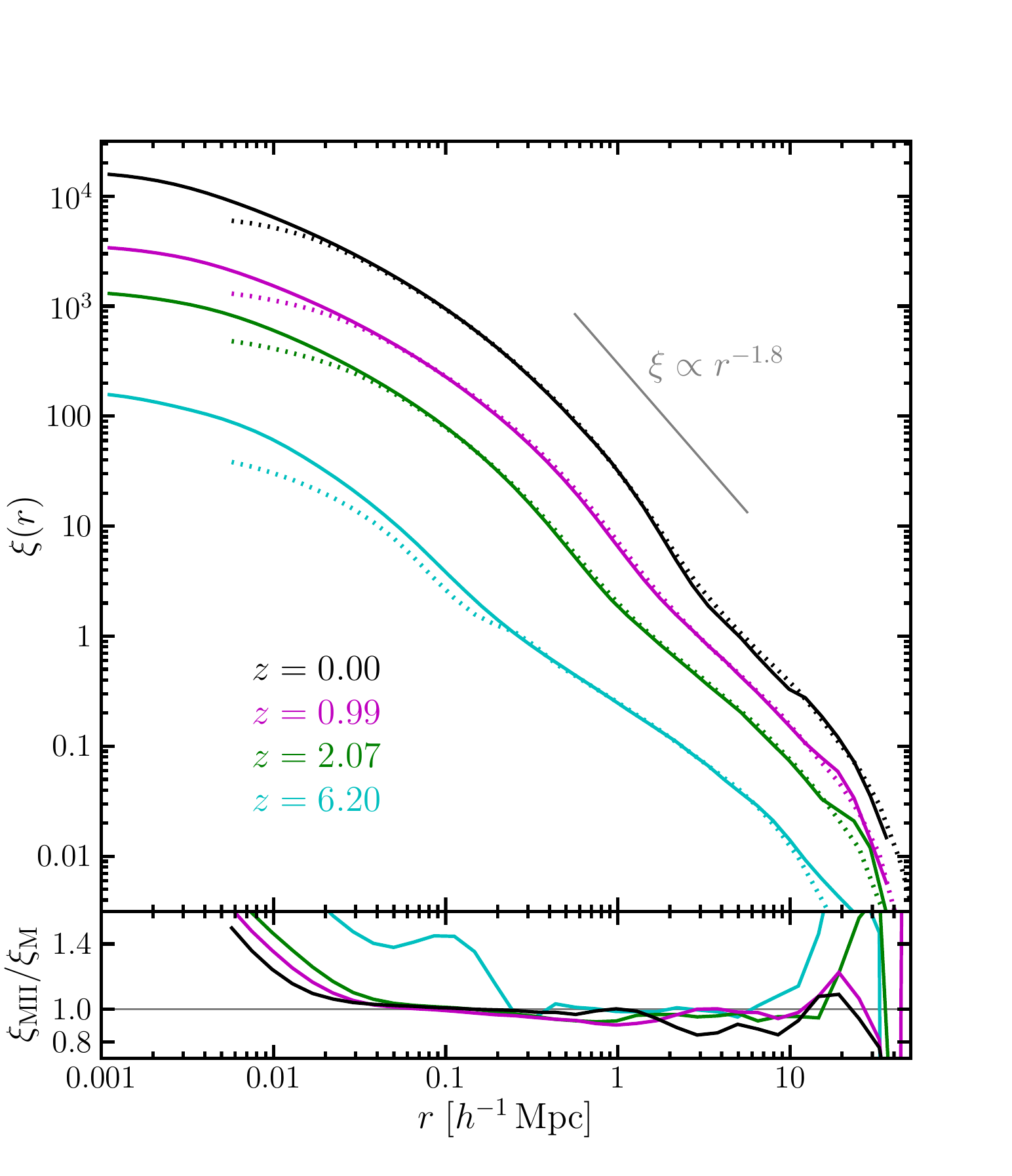}
 \caption{
   \label{fig:correl}
   Measurements of the two point correlation function $\xi$ as a function of
   comoving separation $r$ from the \millen.  We show four redshifts: $z=0.0$
   (black curves), 0.99 (magenta), 2.07 (green), and 6.20 (cyan) and we compare
   with $\xi(r)$ from the MS (dotted curves) at the same redshifts.  On large
   scales, the correlation functions from the two simulations agree quite well.
   On small scales ($\la 0.020 \, \hmpc$ in physical units), the \millen\
   correlation function amplitude is larger, reflecting structures that are not
   resolved in the MS.  The bottom panel focuses on these differences by
   plotting the ratio of the correlation function from the \millen\ to that from
   the MS.  }
\end{figure}
The spatial two-point correlation function of the density field is given by 
\begin{equation}
 \label{eq:correlationFunction}
 \xi(r) = \langle \delta(\mathbf{x}) \delta(\mathbf{x+r}) \rangle \;,
\end{equation}
or equivalently, by a Fourier transform of the power spectrum:
\begin{equation}
 \label{eq:correlationFunction2}
 \xi(r) = \int \Delta^2(k) \frac{\sin{(kr)}}{kr} \, \dd \log k \;.
\end{equation}
The correlation function is a useful measure of the spatial clustering of dark
matter: it gives the excess probability of finding pairs of particles at a
given separation relative to a Poisson distribution.  Figure~\ref{fig:correl}
shows $\xi(r)$ at redshifts 0, 0.99, 2.07, and 6.20, with results from the
MS at the same redshifts also plotted for comparison.  (Note
that the scale on the horizontal axis of Figure~\ref{fig:correl} is in
co-moving units.)  

The correlation function shows a prominent feature at $r\approx 2 \, \hmpc$
(for $z=0$).  This is the well-known transition between the `one-halo' and
`two-halo' contributions: on smaller scales the correlation function is
dominated by dark matter particle pairs within the same halo, while on larger
scales it is dominated by pairs in separate halos \citep{peacock2000,
  seljak2000, ma2000, scoccimarro2001, cooray2002}.  At no redshift is the
correlation function even roughly approximated by a single power law.  This is
in stark contrast to observations of galaxy correlation functions
\citep{zehavi2002, hawkins2003} and the stellar mass autocorrelation function
\citep{li2009} at low redshift, which show a remarkably good power-law behavior
$\xi \propto r^{-1.8}$ over $10^{-2} \la r \la 10 \,\hmpc$ (a gray line with
this relation is also plotted in Figure~\ref{fig:correl} for comparison).

From $z=2$ to $z=0$, the agreement between the \millen\ and MS results is quite
good from $0.03$ to $2 \,\hmpc$.  The \millen\ result lies approximately 10\%
below the MS $\xi(r)$ on scales of 2-10 $\hmpc$; this is the range where the
contributions from pairs in separate halos become important and is likely an
indicator of cosmic variance in the two-halo term.  On small scales, the
\millen\ correlation function lies above that of the MS.  This is a result of
the higher force and mass resolution of the \millen: low-mass halos that were not
resolvable in the MS also boost the clustering on small scales.  The MS
correlation function is noticeably lower in amplitude than $\xi(r)$ from
the \millen\ at $z=6$ for $r \la 0.2 \,\hmpc$.  This coincides with the mean
interparticle spacing, $0.231 \,\hmpc$ comoving for the MS, and is therefore
most likely due to discreteness in the glass-like particle load used for the
initial conditions.

\section{Dark matter halos}
\label{sec:darkMatterHalos}
Understanding how dark matter overdensities grow and ultimately
virialize into highly non-linear structures is an extremely difficult
problem from a theoretical perspective. No rigorous analytic
techniques are available for use in both the linear and non-linear
regimes.  The most successful model for dark matter halo formation
(\citealt{press1974, bond1991, lacey1993, sheth2001}; see
\citealt{zentner2007} for a recent review of extended Press-Schechter
theory) relates the abundance of halos at mass $M$ and redshift $z$ to
the initial power spectrum of density fluctuations and to the
well-understood regime of linear growth. It considers 
the linear overdensity field smoothed using a spherical top-hat filter
(in real space) containing mass $M$ and extrapolated using linear
theory to redshift $z$.  The variance of this smoothed field is:
\begin{equation}
 \label{eq:massVariance}
 \sigma^2(M, z) = d^2(z) \int \Delta_{\rm lin}^2(k) \, W^2(k; M) \, d \log k \;, 
\end{equation}
where $d(z)$ is the linear growth factor at redshift $z$ with
normalization $d(z\!=\!0)\!=\!1$, $\Delta_{\rm lin}^2(k)$ is the
linear power spectrum extrapolated to $z=0$, and $W(k; M)$ is the
Fourier transform of a real-space spherical top-hat filter. Each dark
matter particle is assigned to a halo of mass $M$ at redshift $z$,
where $M$ is taken to be the largest filter scale for which the
smoothed linear overdensity at the particle's position (extrapolated
to redshift $z$) exceeds a threshold value
$\delta_c$.\footnote{$\delta_c=1.68647$ at all redshifts for an
 Einstein-de Sitter universe.  For our $\Lambda$CDM cosmology,
 $\delta_c$ varies slightly, from this standard value at high
 redshift to 1.6737 at redshift zero due to a weak dependence on
 $\Omega_m(z).$} A characteristic halo mass $\mstar$ can then be
defined at each redshift via $\sigma(\mstar, z) = \delta_c$.  In this
model,  the halo multiplicity function $f(\sigma)$, which is related to
the comoving number density of halos via
\begin{equation}
 \label{eq:haloAbundance}
  f(\sigma) =   \frac{M}{\bar{\rho}(z)} \, \frac{\dd \, n(M, z)}{\dd \log \sigma^{-1}}\,,
\end{equation}
takes on a universal form (for perspectives on universality of the halo mass
function, see \citealt{jenkins2001, reed2007, lukic2007, cohn2008, tinker2008}).

\subsection{Mass Function}
\begin{figure}
 \centering
 \includegraphics[scale=0.58, viewport=0 0 421 405, clip]{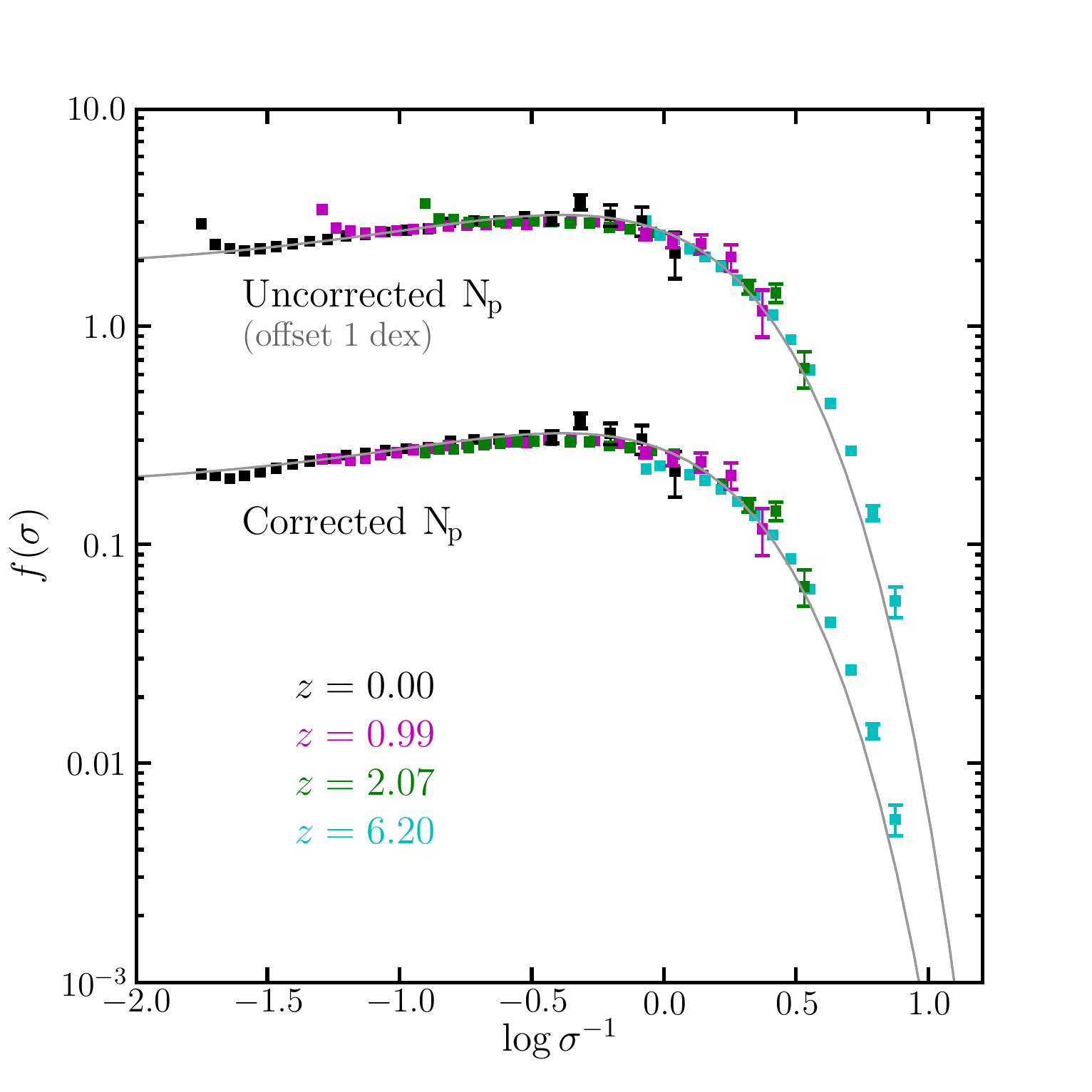}
 \caption{
   \label{fig:massFunc1}
   The halo multiplicity function $f(\sigma)$ as a function of $\sigma(M,
   z)^{-1}$ at four redshifts from the \millen: $z=0$ (black
   squares), 0.99 (magenta), 2.07 (green), and 6.2 (cyan).  We compute the
   multiplicity function both with the Warren et al. $N_p$ correction (lower
   data points) and without the correction (upper points, offset by 1 dex for
   clarity).  Also overplotted is the \citet{warren2006} fit to the halo
   multiplicity function.  The \millen\ multiplicity function shows universal
   behavior when scaling with respect to redshift, with deviations at the 10\%
   level.  Halo masses here are defined to be $M_{\rm FOF}$.}

\end{figure}
\begin{figure}
 \centering
 \includegraphics[scale=0.58, viewport=0 0 421 405, clip]{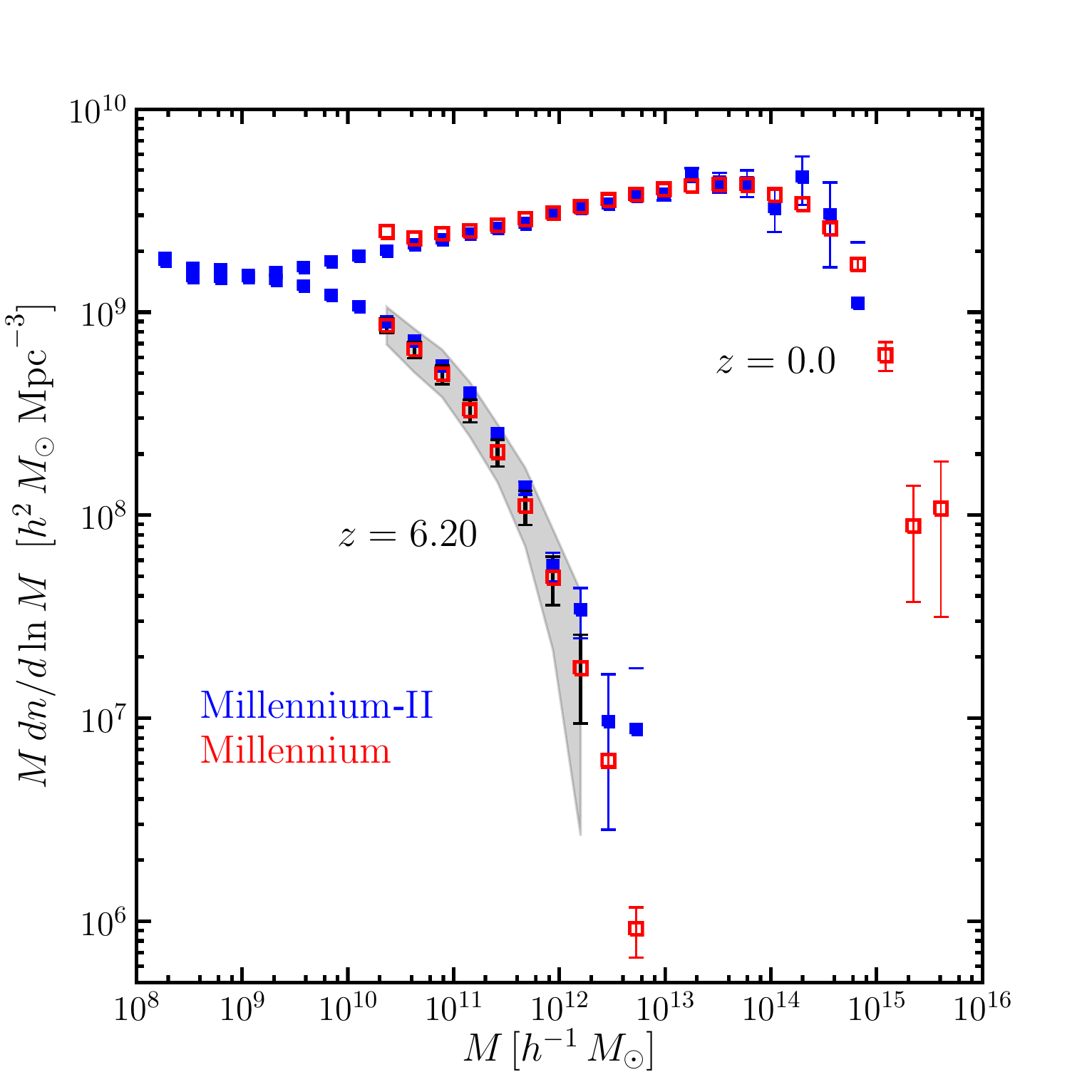}
 \caption{
   \label{fig:massFunc2}
   FOF mass function for the \millen\ (solid blue squares) and for the MS
   (open red squares) compared at redshift 6.2 and 0.  The
   redshift zero mass functions are in excellent agreement over the
   entire range where the two simulations overlap.  At redshift 6.2,
   the \millen\ points lie systematically above those from
   the MS.  The shaded gray region shows the range of mass functions
   obtained from subdividing the MS into 125 cubes with volume equal
   to the \millen\ and computing a mass function for each sub-volume.
   The \millen\ points are well within the scatter, indicating that
   the difference is likely due to the small volume of the \millen. }
\end{figure}
In Figure~\ref{fig:massFunc1} we plot the halo multiplicity function
$f(\sigma)$ from the \millen\ at four redshifts: $z=0.0, \, 0.99, \, 2.07$, and
$6.20$.  Here we define halo mass as $M_{\rm FOF}$; see \citet{hu2003} and
\citet{tinker2008} for discussions of halo multiplicity functions computed
using spherical overdensity masses.  We include Poisson error bars for all bins
containing fewer than 400 halos, as Poisson errors dominate sample variance at
all masses (e.g., \citealt{hu2003}).  Results are plotted both using the
\citet{warren2006} correction\footnote{This correction is $N_p \rightarrow
  N_{\rm corrected}=N_p\,(1-N_p^{-0.6})$} for sampling bias in $N_p$ (lower set
of data points) and without this correction (upper data points, offset upward
by 1 dex for clarity).  We exclude halos that have $N_p< 20$ when using the
corrected $N_p$.  The simulation has fixed mass resolution but $\sigma(M,z)$
evolves significantly with time, so by comparing multiplicity functions at
several redshifts we can probe a large range in $\sigma$.

The multiplicity function within the \millen\ does seem to have a
universal form: where the data overlap, the agreement in $f(\sigma)$ is quite
good.  It thus appears possible to compute the multiplicity function at any
redshift simply by combining the linear growth factor $d(z)$ with the rms
amplitude of fluctuations as a function of mass at redshift zero.  This
agreement is at least as good for the uncorrected points, excluding bins
containing halos with fewer than 100 particles.  (We note, however, that Warren
et al. used a minimum of 400 particles per halo in deriving their fitting
parameters; in this regime, both the corrected and uncorrected points seem to
exhibit `universality.')  The multiplicity function does not agree precisely
with the Warren et al. fit (gray line) in either case; however, the volume of
the \millen\ is not sufficiently large to obtain statistically precise results in
the high $\sigma^{-1}$ regime due to cosmic variance.

Figure~\ref{fig:massFunc2} compares the FOF mass function at redshifts 0 and
6.2 determined from the \millen\ (solid blue squares) with the MS
mass function (open red squares).  Poisson error bars are included for all bins
with fewer than 400 halos and the data points do not include the Warren et
al. correction for the sampling bias in $N_p$.  At $z=0$, the agreement between
the two simulations is excellent for all halo masses (excluding bins containing
halos with fewer than 100 particles).  Combining the two allows for a
consistent measurement of the halo mass function over seven decades in halo
mass.  At $z=6.2$, the \millen\ mass function lies systematically above that of
the MS.  The most likely explanation of this difference is cosmic variance: the
halos probed by either simulation at $z=6.2$ are inherently rare objects, as
the characteristic mass $\mstar$ is $4.5 \times 10^{5}\, \hmsun$ at that
time\footnote{The minimum halo mass in the MS, $10^{10} \,\hmsun$, corresponds
  to a peak height $\nu \equiv \delta_c/\sigma(M, z)$ of 1.5 at $z=6.2$,
  which is equivalent to a mass of $7 \times 10^{13} \,\hmsun$ at $z=0$.}.
Furthermore, the \millen\ probes only 1/125$^{\rm th}$ the volume of
the MS, making statistical fluctuations much more likely.

In order to estimate the effects of cosmic variance on these mass functions, we
divided the MS into 125 disjoint sub-cubes, each with the same volume as the
\millen, and we measured the scatter in mass functions and in the mean matter
densities $\bar{\rho}_m$ computed from these sub-volumes at $z=6.2$.  The full
range of these mass functions is plotted as a gray shaded region in
Figure~\ref{fig:massFunc2}, while the rms values at each mass are shown as
black error bars on the MS data points.  The \millen\ points typically lie
slightly outside of the rms region but well within the full distribution of
mass functions, indicating that they are fully consistent with the MS when the
volume of the \millen\ is taken into account.  We emphasize that the variation
in the mass functions between the 125 MS sub-cubes is {\it not} due to
differences in the mean matter density, as the rms scatter in $\bar{\rho}_m$ is
only 2\% while the rms scatter in the mass function exceeds 8\% (the full range
of the scatter exceeds $\pm 20\%$) for all of the data points.

\subsection{Bias}
\label{sec:bias}
\begin{figure}
 \centering
 \vspace{0.25cm}
 \includegraphics[scale=0.58, viewport=0 0 421 421]{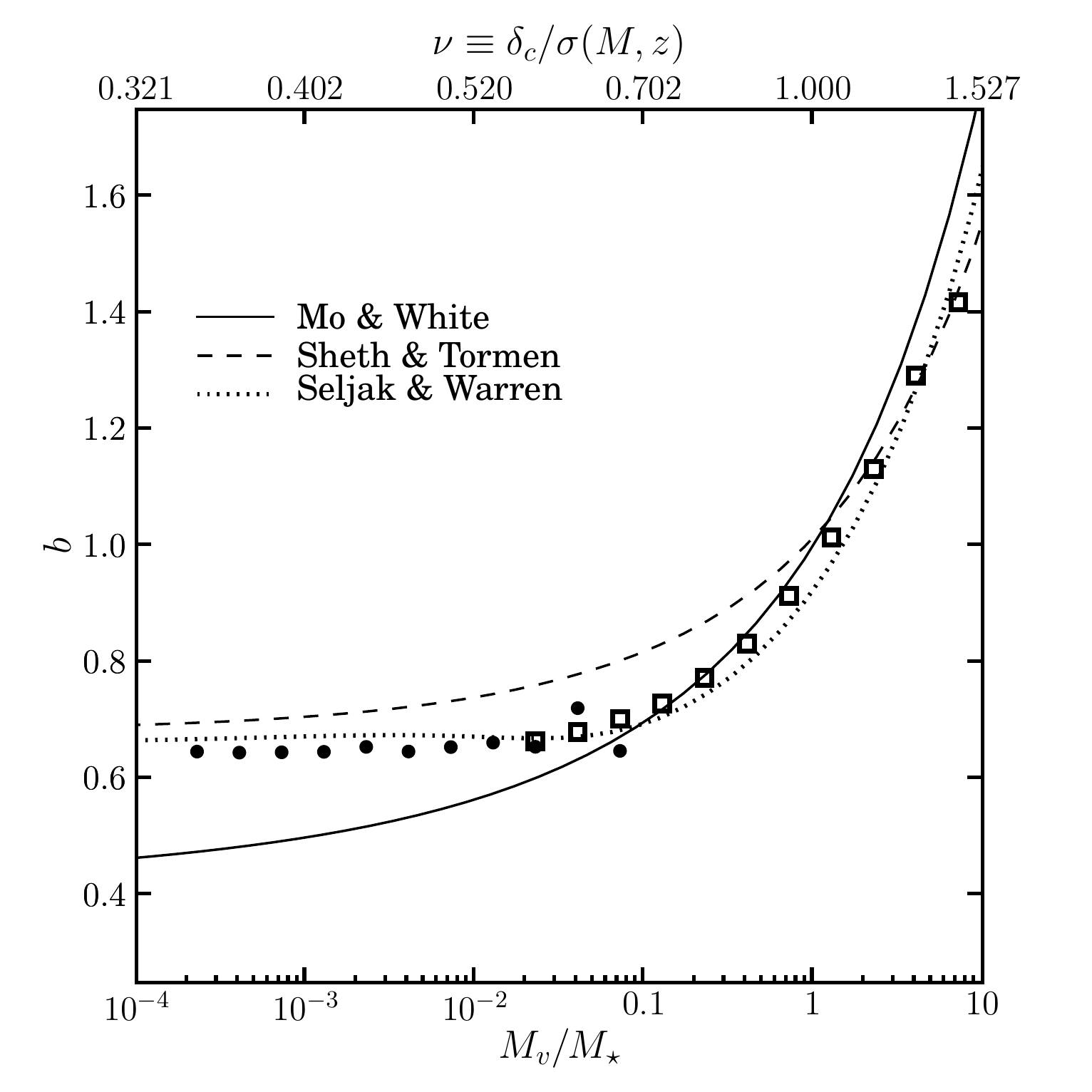}
 \caption{
   \label{fig:bias}
   Halo bias at redshift zero.  We combine results from the \millen\
   (filled circles) and the MS (open squares) to
   explore bias from $10^{-4}$ to $10 \, \mstar$.  As expected, the bias
   decreases as the halo mass decreases, reaching $b(\mstar) \approx 1$.  At
   very low masses ($M_v / \mstar \la 2 \times 10^{-2}$ or $\nu \la 0.55$),
   the bias reaches an asymptotic value of 0.65.  }
\end{figure} 
Dark matter halos do not cluster in the same way as the underlying mass density
field but rather exhibit a bias relative to the dark matter.  \citet{mo1996},
building on the earlier work of \citet{efstathiou1988} and \citet{cole1989},
showed that the two-point correlation function of halos should be simply
related to that of the mass density field. According to the excursion set
model, on large scales one should find that
\begin{equation}
 \label{eq:twopointHalos}
 \xi_{hh}(M, z; \,r)=b^2(M, z)\,\xi_{mm}(z; \,r) \,,
\end{equation}
where the bias factor $b$ is given by 
\begin{equation}
 \label{eq:bias}
 b(M, z) = 1+\delta_c^{-1} \, (\nu^2-1)\,.
\end{equation}
Massive halos ($M \ga \mstar$) are therefore predicted to cluster more strongly
than the underlying mass density field while low-mass halos should cluster less
strongly.  This basic picture has been extensively validated, with newer models
making improved quantitative predictions for bias \citep{jing1998, sheth1999,
  sheth2001, seljak2004}.

The bias of low-mass halos ($M \ll \mstar$) remains an unresolved issue.  In
the Mo \& White model, $b \rightarrow 1-\delta_c^{-1} \approx 0.4$ for $M \ll
\mstar$.  \citet{sheth1999} and \citet{seljak2004}, on the other hand, find $b
\approx 0.7$ in this same regime.  The mass resolution of the \millen\ allows us to
study bias for $M \ll \mstar$: halos with 200 particles, whose number densities
and spatial distributions are certainly well-resolved, correspond to $2 \times
10^{-4} \, \mstar$ at $z=0$.

Halo bias at redshift zero is shown in Figure~\ref{fig:bias} for masses down to
$2 \times 10^{-4}\, \mstar$.  The bias for $M_v \ll \mstar$ is constant at $b
\approx 0.65$ over approximately two decades in mass, from $M_v/\mstar=2\times
10^{-4}$ to $2 \times 10^{-2}$.  In terms of peak height $\nu$ (shown on the
upper horizontal axis of Figure~\ref{fig:bias}), the bias is constant for $\nu
< 0.55$.  Above $0.1 \, \mstar$, the bias rises rapidly with mass, reaching
$b=1$ at $M$ slightly greater than $\mstar$ and $b \approx 1.5$ for $10\,
\mstar$.

Figure~\ref{fig:bias} also shows the predictions for $b(M)$ from three fitting
formulae: the original Mo \& White prediction (solid curve), the Sheth \&
Tormen model (dashed), and the fit from Seljak \& Warren (their eq. 5; dotted).
The Seljak \& Warren fit clearly agrees best with the data over the range
plotted, though it slightly underpredicts the bias at $M > \mstar$ and slightly
overpredicts it at $M < \mstar$.  These differences are only at the 5\% level,
however.

\section{Halo Formation}
\label{sec:haloFormation}
\subsection{Formation Times}
\begin{figure}
 \centering
 \includegraphics[scale=0.58, viewport=0 0 421 405, clip]{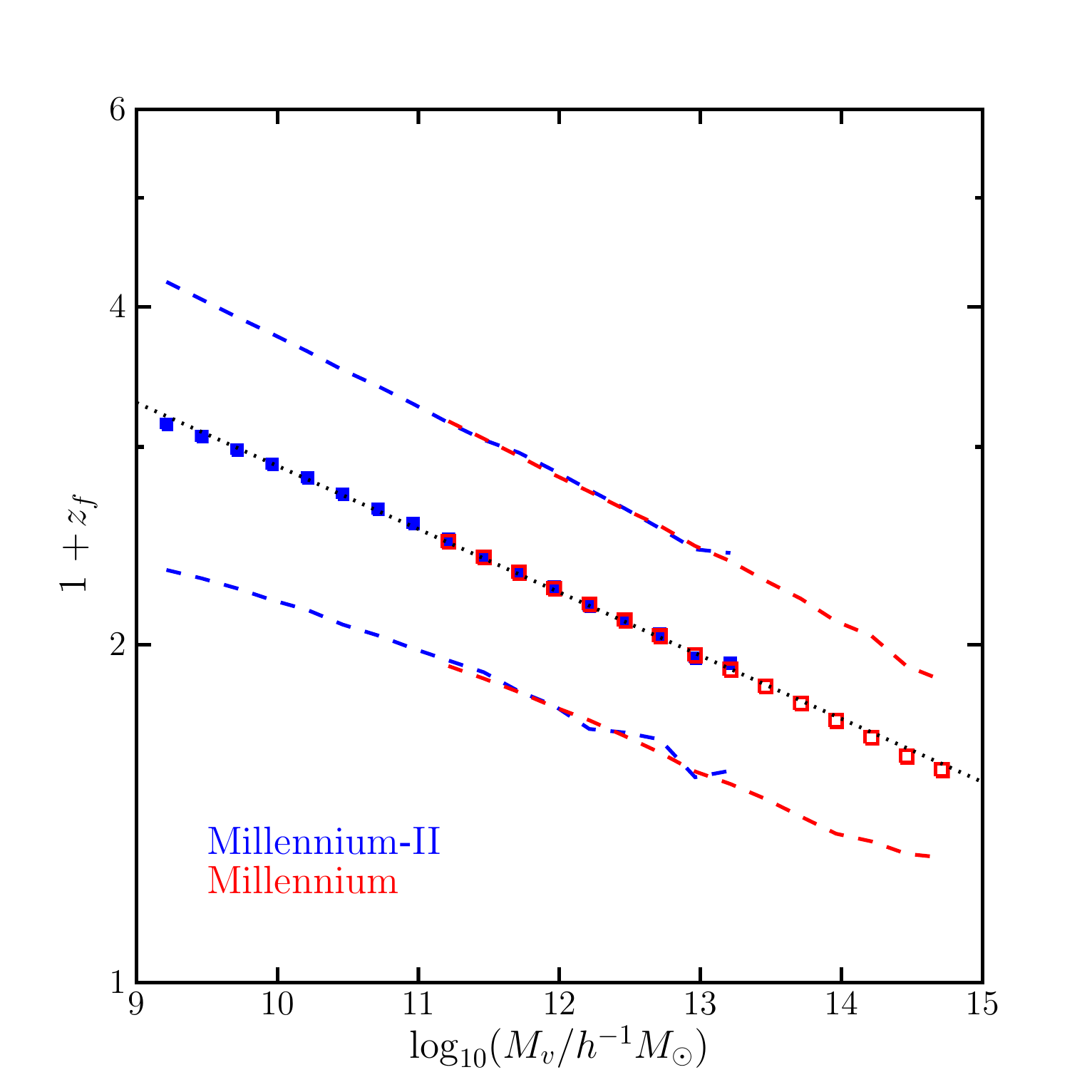}
\caption{
   \label{fig:zform}
   Mean half-mass formation redshift as a function of mass for halos from the
   \millen\ (filled blue squares) and the MS (open red squares).  We also show
   the relation for the 16\% earliest and latest forming halos (dashed lines)
   and the best-fitting linear relation between $\log (1+z_f)$ and $\log M_v$
   (black dotted line; see Equation~(\ref{eq:zform}) for the fitting
   parameters). This fit deviates from the mean relation by less than 2\% over
   the entire range of masses plotted.}
\end{figure}

The hierarchical nature of $\Lambda$CDM models means that the typical formation
redshift $z_f$ of halos with mass $M$ is a decreasing function of $M$.  The
form of the relation between $z_f$ and $M$ and its intrinsic scatter are
important for a number of applications, such as understanding how well galaxy
properties can be predicted by halo mass alone, and how well a halo's history
can be predicted from its present-day properties.  Such characterizations are
complicated by the fact that the most useful definition of formation time
depends on the question one is asking\footnote{See, e.g., \citealt{li2008} for
  several possible definitions of $z_f$}.  For example, the innermost region of
a halo -- where the galaxy resides -- typically assembles much earlier than the
outer regions which contain most of the mass \citep{zhao2003, gao2004}.

One of the simplest definitions of formation redshift is the time at which a
halo's main progenitor reaches a fixed fraction of its present-day mass.  We
use the merger trees described in Section~\ref{sec:mergerTrees} to trace each
FOF halo back in time and define its formation redshift $z_f$ as the first
redshift at which one of the halo's progenitors reached half of the halo's
redshift zero mass (we interpolate between snapshot redshifts to obtain $z_f$).
This `half-mass' formation time is the most common choice of formation time in
the literature. We use $M_v$ as the definition of halo and progenitor
mass when estimating such formation times (we have checked that the following
results are insensitive to halo mass definition).

In Figure~\ref{fig:zform}, we show the mean relation between halo mass $M_v$
and formation time $z_{f}$.  In order to determine what mass is required for
converged results, we compute $z_{f}$ from the MS and compare with the \millen.
We find that the two simulations are in excellent agreement above a redshift
zero mass of $10^{11} \, \hmsun$, corresponding to approximately
150 particles in the MS.  This is the convergence limit we adopt, so we
consider all halos with masses greater than $10^{9}\, \hmsun$ in the \millen.
We only include halos that {\tt SUBFIND} determines to be bound, although in
practice, this restriction makes almost no difference as the fraction of
subhalos with $N_p > 100$ that are unbound is very small.  Over the entire
range where halo formation can be resolved ($9 \le \log_{10}(M_v/ \hmsun) \le
14.7$), a simple linear fit in $\log(1+z_{f})$ versus $\log(M_v)$ provides an
excellent description of the data:
\begin{equation}
 \label{eq:zform}
 1+z_f = 2.89 \, \left(\frac{M_v}{10^{10} \,\hmsun}\right)^{-0.0563} \,.
\end{equation}
The maximum deviation between the binned data and Equation~(\ref{eq:zform}) is
1.8\% over the entire region where: (1) there are at least 100 halos per bin;
and (2) halos have at least 125 particles.  The 1-$\sigma$ scatter in the
relation, defined as the logarithmic difference between the $84^{\rm th}$ and
$16^{\rm th}$ percentiles of the data (which are shown in dashed lines in
Figure~\ref{fig:zform}), decreases gradually from $\sigma_{\log 1+z}=0.6$ to
0.4 as $M_v$ increases from $10^9$ to $5 \times 10^{14}\,\hmsun$.

\citet{neto2007} previously studied the relation between median $z_f$ and halo
mass for a set of relaxed halos from the MS.  Their fit is similar to ours
although the parameters differ slightly (their exponent is -0.046 and their
normalization is approximately 2.74 using our form of the fitting formula) due
presumably to the selection criteria used for the halos they studied, their use
of $\mtwo$ rather than $M_v$, and the difference between the median and the
mean relation.  \citet{mcbride2009} have also recently computed the formation
times of massive halos from the MS and fitted to the relation $z_f = a
\log_{10}(M/10^{12} \,M_{\odot}) + b$.  This gives very similar results to ours
over the range of the data they used ($M \ga 10^{12} \, \hmsun$), with
differences at the 5\% level after an empirical normalization correction due to
a different mass definition. Their formula underestimates $z_f$ from the \millen\
at lower masses: at $10^{9} \, \hmsun$, the difference is approximately 20\%.

\subsection{Clustering and Formation Times}
\begin{figure}
 \centering
 \includegraphics[scale=0.58, viewport=0 0 421 405, clip]{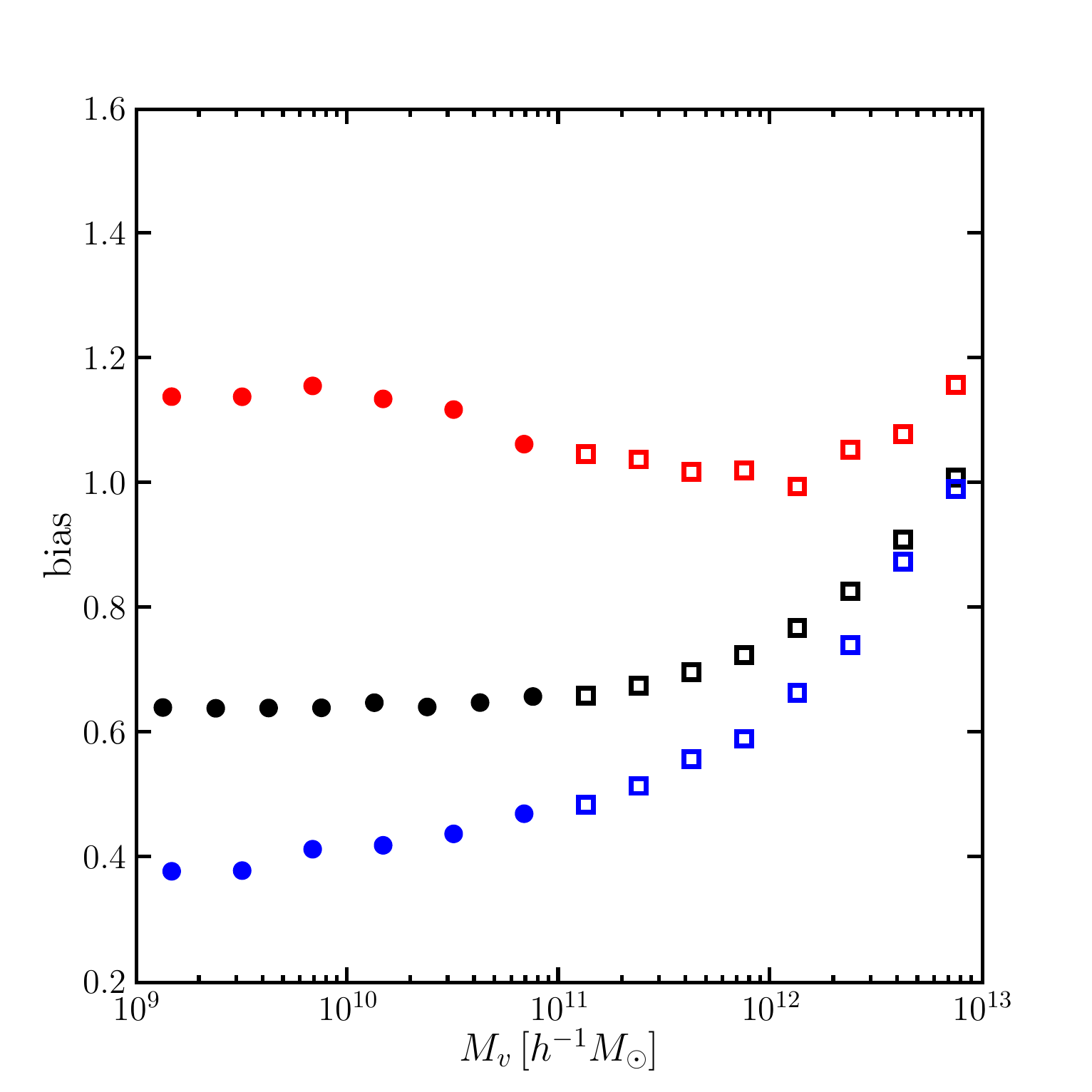}
 \caption{
   \label{fig:assemblyBias}
   Assembly bias from the \millen\ (open symbols) and the MS (filled
   symbols).  We plot the bias of the oldest 20\% of halos (red
   points) and the youngest 20\% of halos (blue points), as well as
   the bias of the full halo sample (black points), as a function of
   halo mass.  The oldest halos cluster much more strongly than the
   youngest, a ratio of 2.85 in the bias (which is over a factor of 8
   in correlation amplitude).  At low masses, the bias for each
   subset reaches an asymptotic value ($\approx 0.4$ for the young
   halos, $\approx 1.15$ for the old halos).}
\end{figure}
In the simple version of the excursion set model of structure formation, which
is based on top-hat $k$-space filtering, halo formation is governed by Markov
random walks of the (linearly-extrapolated) mass density field $\delta({\bf x};
M)$.  A direct consequence of this aspect of excursion set theory is the
prediction that properties such as the clustering of halos depend on halo mass
alone and not on halo assembly history.  Recent $N$-body simulations have
produced results that contradict this prediction, however.  \citet{gao2005}
used the MS to show that clustering depends strongly on formation time at
masses $M \la \mstar$: they found that early-forming halos cluster much more
strongly than late-forming halos, indicating an ``assembly bias''.  Subsequent
work confirmed these findings and extended the results to the $M \ga \mstar$
regime and to halo properties other than formation time, including
concentration, substructure content, and spin (e.g., \citealt{harker2006,
  wechsler2006, wetzel2007, jing2007, bett2007, gao2007, dalal2008, angulo2008a})

With the \millen, we are able to investigate assembly bias at much
lower masses than was previously possible: $M \approx 10^{-4}\, \mstar$ or
equivalently $\nu \approx 0.32$.  We split each mass bin into the oldest and
youngest 20\% of halos and compute the bias factor $b(M)$ in the same manner
described in Section~\ref{sec:bias}.  Our results for the dependence of
clustering on formation time are presented in Figure~\ref{fig:assemblyBias}.
We plot the bias of halos as a function of mass at redshift zero for the entire
sample of halos (black symbols) as well as for the oldest 20\% (red symbols)
and youngest 20\% (blue symbols) of halos at each mass.  We show results for
the \millen\ (filled circles) for $M \la 2 \times 10^{11} \, \hmsun$ and for
the MS (open squares) at higher masses in order to maximize statistical
significance.

Over virtually the full range of masses probed by the \millen, the biases of
the oldest and youngest halos are approximately constant.  The bias of young
halos is substantially lower than that of old halos, however, in agreement with
previous work.  The oldest 20\% of halos at a given mass have a slight positive
bias with respect to the dark matter distribution.  The youngest 20\% have a
bias that is approximately $b=0.4$; interestingly, and perhaps coincidentally,
this is the value of $1-\delta_c^{-1} \approx 0.4$ predicted in the $\nu \ll 1$
regime by Press-Schechter and excursion set models (\citealt{cole1989, mo1996};
see also \citealt{dalal2008}).

\citet{li2008} have suggested that the magnitude of assembly bias depends on
the adopted definition of formation time.  To investigate whether our
definition of formation redshift, $M(z_f)=\frac{1}{2} \, M(z=0)$, influences
our results, we have repeated our analysis with a new definition: $M(z_f) =
\frac{1}{4} \, M(z=0)$.  We have also tested whether computing the half-mass
formation time relative to $M_{\rm FOF}$ rather than to $M_v$ affects our results.
Neither of these changes has any detectable influence, so the results we obtain
using the standard half-mass formation time appear robust.

\section{`Milky Way' Halos in Millennium-II}
\label{sec:aquarius}
\begin{table}
\begin{tabular}{lcccc}
\hline
\hline
{\bf Halo} &  
$\mathbf{\mtwom}$ &  
$\mathrm{\mathbf{\vmax}}$ &	
$\mathbf{R_{max}}$&
$\mathrm{\mathbf{V_{max, \, sub}}}$\\
& [$\hmsun$] & [km/s] & [$\hkpc$] & [km/s] \\
\hline
\hline
Aq-A &  $1.842 \times 10^{12}$ &  208.49 &  20.54 &  60.42\\
MS-II-A &  $1.826 \times 10^{12}$ &  210.20 &  21.87 &  61.68\\
\hline
Aq-B &  $7.629 \times 10^{11}$ &  157.68 &  29.31 &  48.31 \\
MS-II-B &  $7.470 \times 10^{11}$ &  156.21 &  27.98 &  45.87 \\
\hline
Aq-C &  $1.641 \times 10^{12}$ &  222.40 &  23.70 &  87.07\\
MS-II-C &  $1.682 \times 10^{12}$ &  222.03 &  23.88 &  89.70\\
\hline
Aq-D &  $1.839 \times 10^{12}$ &  203.20 &  39.48 &  90.63\\
MS-II-D &  $1.944 \times 10^{12}$ &  204.39 &  42.28 &  91.43\\
\hline
Aq-E &  $1.130 \times 10^{12}$ &  179.00 &  40.52 &  40.87\\
MS-II-E &  $1.187 \times 10^{12}$ &  184.15 &  43.00 &  45.36\\
\hline
Aq-F &  $1.107 \times 10^{12}$ &  169.08 &  31.15 &  78.33\\
MS-II-F &  $1.152 \times 10^{12}$ &  167.18 &  34.47 &  81.53\\
\hline
\hline
\end{tabular}
\caption{  
  Comparison of Aquarius level 2 halos (Aq-X) and Aquarius halos in the
  \millen\ (MS-II-X).  $\mtwom$ is the mass of the dark matter halo, $\vmax$ 
  is the maximum value of the circular velocity curve, $\rmax$ is
  the radius at which the circular velocity curve attains its maximum, and
  $V_{\rm max, \, sub}$ is the circular velocity curve maximum for the largest
  subhalo in each halo; all values are quoted at $z=0$. 
\label{table:aquariusComparison}}
\end{table}
The Milky Way can provide us with a unique variety of insights into galaxy
formation, so it is extremely interesting to study the formation of Milky
Way-mass halos for comparison with available and forthcoming observations of
the detailed structure of our Galaxy.  Cosmological simulations of
representative volumes of the universe can provide large, statistically complete
samples of Milky Way-mass halos, but they cannot resolve the full range of
observable structures within the Milky Way.  Even at the mass resolution of
the \millen, for example, the halos of dwarf spheroidal galaxies are just barely
resolvable.  An alternative tack is to focus all of one's computational
resources on the formation of a single galaxy-mass halo, thus allowing
substantially enhanced mass resolution \citep{diemand2007, diemand2008,
  springel2008, stadel2008}.  With this method, one sacrifices statistical
understanding of a representative sample for detailed analysis of one or a few
objects.  Here we discuss how these two approaches may be combined to extract
maximum information about the formation and evolution of galaxy-scale dark
matter halos.

\subsection{The Aquarius Project and Millennium-II}
The Aquarius project \citep{springel2008} is a suite of cosmological
simulations of the formation of Milky Way-mass dark matter halos.  Six halos
(denoted `Aq-A' through `Aq-F') were simulated at up to five different levels
of mass resolution (levels 5 through 1).  The highest resolution simulation,
Aq-A-1, used a particle mass of $1.25 \times 10^3 \, \hmsun$, resulting in
approximately 1.5 billion particles within $\rtwom$ at $z=0$.  The halos
resimulated in the Aquarius project were selected from the cosmological
simulation `hMS' \citep{gao2008}, which followed $900^3$ particles in a $100
\,\hmpc$ box.  Both the cosmological parameters of hMS and the amplitudes and
phases of modes used to generate its initial conditions are identical to those
in the \millen.  As a result, all the structures present in hMS are also
present in the \millen, but with a mass resolution that is a factor of 13.8
times better.  Since the Aquarius halos were selected from hMS, they are also
present in the \millen\ and we can compare their properties in the \millen\ and
in the much higher resolution Aquarius resimulations.

\subsection{Comparing Aquarius and Millennium-II halos}
\begin{figure}
 \centering
 \includegraphics[scale=0.58, viewport=0 0 421 405, clip]{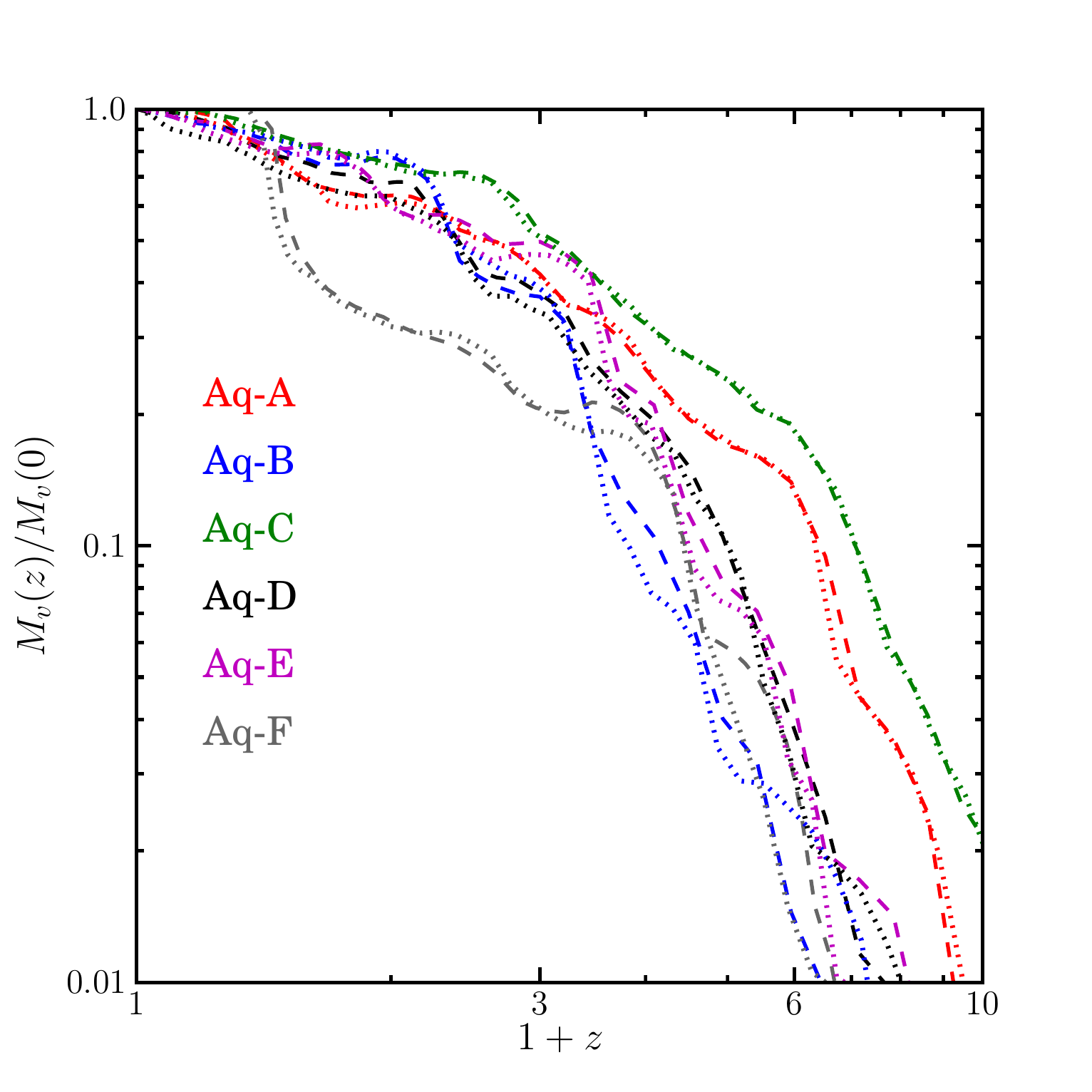}
 \caption{
   \label{fig:mah_aq}
   Mass accretion histories for the six Aquarius level 2 resolution halos
   (dashed lines) and for the corresponding \millen\ halos (dotted lines).
   Even though the mass resolution of the Aquarius simulations is one
   thousand times better than that of the \millen, the individual mass accretion
   histories of the Aquarius halos are captured quite well in the \millen.}
\end{figure} 
As a first test, we compare some basic properties of the Aquarius halos in the
Aquarius level 2 simulations -- where the particle mass is $(0.5-1) \times 10^4
\,\hmsun$ -- and in the \millen, where the mass resolution is approximately
1000 times lower.  Table~\ref{table:aquariusComparison} contains such a
comparison, with halos from the Aquarius simulations in the upper of each set of
two rows (labeled `Aq-X') and halos from the \millen\ in the lower of each set
of rows (labeled `MS-II-X').  The first data column compares $\mtwom$ values
for the halos, showing that their masses agree very well: each $\mtwom$
measured from the \millen\ agrees with the corresponding Aquarius resimulation
to better than 6\% and, for 3 of the 6, to better than 3\%.  The measured
$\vmax$ values all agree within 3\% and, for 5 of the 6, to better than 1.5\%.
The radius at which the circular velocity curve peaks, $\rmax$, typically
agrees within 5\%, with a maximum deviation of 10\%.
Table~\ref{table:aquariusComparison} also lists the maximum circular velocity
of the largest subhalo in each simulation.  In all cases this is the {\it same}
subhalo in the \millen\ and the Aquarius resimulation, and the circular
velocities typically agree to within a few percent.

A more stringent test is to consider the mass accretion history\footnote{for a
  discussion of the statistical properties of mass accretion histories, see
  \citet{lacey1993, wechsler2002, van-den-bosch2002, zhao2003, zhao2003a,
    zhao2008, mcbride2009}.}  for each Aquarius halo and to compare results
from the Aquarius resimulations and from the \millen.  For the \millen\
Aquarius halos, we use the merger trees described in
Section~\ref{sec:mergerTrees} to determine the main progenitor of each halo at
each redshift, and we define the mass accretion history of a halo as the mass
of its main progenitor $M(z)$ at each redshift $z$.  Merger trees have also
been built for the halos in the Aquarius resimulations, and we use these in an
identical manner to define the corresponding mass accretion histories.

The results are compared in Figure~\ref{fig:mah_aq}, where we plot mass
accretion histories over the redshift range $0 \le z \le 9$. We use dashed
lines for the Aquarius resimulations and dotted lines for the \millen.  The
assembly histories of all the halos are reproduced remarkably well at \millen\
resolution.  This is a highly non-trivial test, as the mass resolution of the
Aquarius level 2 simulations is one thousand times better than that of the
\millen: at $z=0$ the Aquarius level 2 halos have on the order of $1.80 \times
10^8$ particles within $\rtwom$ while their \millen\ counterparts have
approximately $1.80 \times 10^5$ particles within this radius.

This agreement is a testament to the integration accuracy of the {\tt GADGET-3}
code and shows that the properties of Milky Way-mass halos and their most
massive subhalos are well converged in the \millen.  As a result, the \millen\ will
be very useful for understanding the statistical properties of Milky Way-mass
halos since it contains over six thousand halos with $11.5 < \log_{10}(M_v/
\hmsun) < 12.5$.  These halos can be used for a detailed statistical
investigation of the growth, internal structure and subhalo populations of
halos similar in mass to that of our own Galaxy (Boylan-Kolchin 2009, in
preparation).  This will test the extent to which the six Aquarius halos are
representative of the full halo population at this mass, thereby allowing
results obtained from the ultra-high resolution Aquarius resimulations to be
interpreted in their full cosmological context.

\section{Discussion}
\label{sec:discuss}
Understanding galaxy formation in a $\Lambda$CDM universe requires knowledge of
physical processes over a wide range of scales, from sub-galactic to
cosmological.  Simulations with volumes large enough to probe the statistics of
large-scale structure and resolution high enough to resolve subhalo dynamics
within galaxy halos are thus critical for this quest.  We have presented
initial results from a new simulation, the Millennium-II Simulation, that can
resolve all halos down to mass scales comparable of the halos of Local Group
dwarf spheroidal galaxies.

Furthermore, the Millennium-II Simulation is closely connected to two other
very large computational endeavors, the Millennium Simulation and the Aquarius
Project.  Throughout this paper we have shown that \millen\ results agree
extremely well with those from the MS, so we can combine the two simulations to
cover an even broader range of physical scales.  We have also shown that the
properties of the resimulated Aquarius halos agree precisely and in
considerable detail with those of their counterparts in the \millen, even though
the resimulations have 1000 times better mass resolution.  This gives us
confidence not only that the assembly histories of Milky Way-mass halos are
well resolved in the \millen, but also that the properties of their more massive
subhalos are converged.  As a result, we will be able to use the \millen\ to make
statistical statements about an ensemble of galaxy-scale halos where the
Aquarius Project provides much higher resolution results for a limited but well
understood subset.

There is much to do with the \millen, both from a dark matter perspective and
from the point of view of galaxy formation.  There are four cluster-size halos
in the \millen\ with over 60 million particles at $z=0$; investigating density
profiles and substructure abundances for these objects, and comparing with
state-of-the-art galaxy-scale simulations such as Aquarius, will shed light on
whether dark matter structures are self-similar as a function of scale.
Subhalo survival times and merger rates, which are crucial ingredients in
galaxy formation models, can be checked in untested regimes.  Furthermore,
directly resolving much smaller halos means that semi-analytic models initially
developed for the Millennium Simulation \citep{springel2005b, croton2006,
  bower2006, de-lucia2007} can now be updated and extended to much lower galaxy
masses (Guo et al. 2009, in preparation).  With this paper, we publicly release
the FOF halo and subhalo catalogs and the subhalo merger trees in a searchable
database\footnote{see http://www.mpa-garching.mpg.de/galform/millennium-II} structured in the same way as has already been done for the Millennium
Simulation \citep{lemson2006}.  As work
progresses, we plan to make \millen\ semi-analytic galaxy catalogs available as
well.  This will allow others to use the \millen\ data for their own research
purposes.

\vskip2cm
\section*{Acknowledgments}
GL works for the German Astrophysical Virtual Observatory (GAVO), which is
supported by a grant from the German Federal Ministry of Education and Research
(BMBF) under contract 05 AC6VHA.  The Millennium-II Simulation was carried out
at the Computing Center of the Max Planck Society in Garching, Germany.  The
Millennium Simulation databases used in this paper and the web application
providing online access to them were constructed as part of the activities of
the German Astrophysical Virtual Observatory.  This work made extensive use of
NASA's Astrophysics Data System and of the astro-ph archive at arXiv.org.

\bibliography{draft}
\label{lastpage}
\end{document}